\documentclass[12pt,letter]{article}
\usepackage{amssymb,amsmath,epsf,graphicx,cite}
\input epsf.sty

\numberwithin{equation}{section}
\numberwithin{table}{section}

\def \ad {\text{ad}}

\begin{document}
{}~ \hfill\vbox{\hbox{UK-08/02} }\break
\vskip 4.1cm

\centerline{\Large \bf The closed string tadpole in open string field theory}
\vspace*{10.0ex}

\centerline{\large \rm Ian Ellwood}

\vspace*{8.0ex}

\centerline{\large \it Department of Physics and Astronomy, }
\centerline{\large \it University of Kentucky, Lexington, KY 40506, USA}
\vspace*{2.0ex}
\centerline{E-mail: {\tt iellwood@pa.uky.edu}}

\vspace*{6.0ex}

\vspace*{6.0ex}

\centerline{\bf Abstract}
\bigskip

We compute a class of gauge invariant observables for marginal
solutions and the tachyon vacuum.  In each case we find that the
observables are related in a simple way to the closed-string tadpole
on a disk with appropriate boundary conditions. We give a sketch of an
argument that this result should hold in general using the BRST
invariance of the closed string two-point function.  Finally, we
discuss the analogous set of invariants in the Berkovits superstring
field theory.

\baselineskip=16pt
\parskip = 3pt

\newpage
\tableofcontents

\section{Introduction}
\label{s:intro}

Open string field theory originated as an attempt to find a classical
theory which, upon quantization, would reproduce the complete
perturbation expansion of open string scattering diagrams
\cite{Witten:1985cc,Giddings:1986wp,Giddings:1986bp,Thorn:1988hm,Zwiebach:1990az}.
Recently, it has become clear that even before quantization, 
classical string field theory contains a rich amount of
information about D-brane physics and, more generally, boundary
conformal field theories
\cite{Sen:1999nx,Zwiebach:2001nj,Taylor:2003gn,Schnabl:2005gv,Okawa:2006vm,Fuchs:2006hw,Ellwood:2006ba,Schnabl:2007az,Kiermaier:2007ba,Okawa:2007ri,Erler:2007rh,Okawa:2007it,Fuchs:2007yy,Fuchs:2007gw,Ellwood:2007xr,Kiermaier:2007vu,Kiermaier:2007ki,Erler:2007xt,Kwon:2008ap,Hellerman:2008wp}.

Indeed, there is a assumption among many string field theory
practitioners that, given a solution of the classical equations of
motion of string field theory $\Psi$, there is a corresponding
boundary $\text{CFT}_\Psi$.  Furthermore, given a boundary
$\text{CFT}_\Psi$ (which is in some unspecified sense ``not too far
away'' from the boundary $\text{CFT}_0$ around which the string field
theory was defined), there is a classical solution $\Psi$ which
shifts us from $\text{CFT}_0$ to $\text{CFT}_\Psi$.

This would-be duality between string fields and boundary CFTs is
obfuscated by the large amount of gauge symmetry in open string field
theory.  For example, if we are working in bosonic cubic string field
theory, and the string field $\Psi$ represents some boundary CFT,
then the string field,
\begin{equation}
  \Psi' = e^{\Lambda} (\Psi + Q_B) e^{-\Lambda} \ ,
\end{equation}
should represent the same boundary CFT for any ghost number 0 string
field, $\Lambda$.  This gauge symmetry has no analogue in boundary
conformal field theory so, if we wish to compare the two sides of the
duality, it is useful to consider gauge-invariant quantities.

The list of known gauge-invariant objects is very short.  For the
bosonic string, one has the classical action \cite{Witten:1985cc}, 
\begin{equation} \label{HIinvariant}
   S(\Psi) = \frac{1}{2} \int \Psi * Q_B \Psi + \frac{1}{3} \int \Psi*\Psi*\Psi \ ,
\end{equation} 
and the quantities discovered independently by Hashimoto and Itzhaki \cite{Hashimoto:2001sm} and Gaiotto, Rastelli, Sen, and Zwiebach \cite{Gaiotto:2001ji}, which take the
form\footnote{These invariants were first
introduced in a different context by Shapiro and Thorn in
\cite{Shapiro:1987gq,Shapiro:1987ac}.},
\begin{equation} \label{InvariantIntro}
  W(\Psi,\mathcal{V}) = \langle \mathcal{I}| \mathcal{V}(i) |\Psi\rangle \ ,
\end{equation}
where $\mathcal{I}$ is the identity string field, and $\mathcal{V} = c
\bar c \mathcal{O}^{\text{m}}$ is an on-shell closed-string vertex
operator inserted at the midpoint of the string (which is at the point
$z=i$ in the standard UHP coordinates).  

While the classical action has a straightforward interpretation, it is
less clear what the invariants (\ref{InvariantIntro}) compute.  In fact,
since $W(\Psi,\mathcal{V})$ involves the identity field, one might
worry that it would be singular, but, as we'll see in explicit
computations, it is well-defined for the known solutions.  

Since $W(\Psi,\mathcal{V})$ is gauge-invariant, it should correspond
to some definite quantity in the CFT associated with $\Psi$.  In this paper we
motivate the following proposal:

\bigskip
\noindent
Let the string field theory of interest be defined
around a boundary CFT${}_0$.  Let $\Psi$ be a string-field associated
to the boundary $\text{CFT}_\Psi$.  Then
\begin{equation} \label{MainRelation}
   W(\Psi, \mathcal{V}) = \mathcal{A}^{\text{disk}}_\Psi(\mathcal{V})- \mathcal{A}^{\text{disk}}_0(\mathcal{V})  \ ,
\end{equation}
where $\mathcal{A}^{\text{disk}}_\Phi(\mathcal{V})$ is the disk
amplitude with one closed string vertex operator $\mathcal{V}$ and
boundary conditions given by CFT${}_\Phi$.

As we will show, this relationship can be derived from the BRST
invariance of the closed string two-point function.  This derivation
is very delicate both in its use of BRST invariance and its implicit reliance
 on certain assumptions about
the nature of the string fields used in the computation of the
invariants.  As such, our derivation is non-rigorous, and we consider the
fact that (\ref{MainRelation}) holds in explicit examples as important
evidence that it is correct.

The relation (\ref{MainRelation}) can be viewed in two ways: First,
given a $\Psi$, we may compute the left hand side for all possible
$\mathcal{V}$ to determine the complete {\em physical} part of the boundary
state of $\text{CFT}_\Psi$.  Second, given a boundary state of some
boundary CFT for which we don't know the associated $\Psi$, we can use
(\ref{MainRelation}) to find a  number of linear constraints on
$\Psi$.  These may aid in the search for new solutions to the string
field theory equations of motion, though it does not seem that they
are enough information to derive a string field theory solution given
a CFT since the on-shell condition on the closed string field puts tight
restrictions on its form in most cases.

Having given an interpretation for $W(\Psi,\mathcal{V})$, it is
natural to extend the construction to the Berkovits open superstring
field theory \cite{Berkovits:1995ab,Berkovits:1998bt,Berkovits:2000zj}.  
The string field in this case has a different gauge invariance,
\begin{equation}
  e^{\Phi} \to e^{Q_B \Lambda} e^{\Phi} e^{\eta_0 \Lambda'}
\end{equation}
where $\Lambda$ and $\Lambda'$ are independent gauge parameters and
$\eta_0$ is the zero mode of $\eta$ in the $\eta$, $\xi$, $\phi$
superconformal ghost system.  Nonetheless, a set of invariants, which are very similar to
the bosonic invariants was written down in \cite{Michishita:2004rx}.

We use a slightly different, but equivalent, form of these invariants: As has held true
in a number of examples \cite{Erler:2007rh,Okawa:2007ri,Okawa:2007it,Kiermaier:2007ki}, the analogue of the bosonic string field
$\Psi$ in the superstring is $e^{-\Phi} Q_B e^{\Phi}$.  This leads to
a set of invariants in superstring field theory,
\begin{equation}
  \widehat{W}(\Phi,\mathcal{V}) = \langle \mathcal{I}| \mathcal{V}(i)| e^{-\Phi} 
  Q_B e^{\Phi}\rangle \ ,
\end{equation}
where $\mathcal{V}$ is a weight zero primary field inserted at the
midpoint which satisfies
\begin{equation}
  Q_B \eta_0 \mathcal{V} = 0 \ .
\end{equation}
The operator $\mathcal{V}$ lives in the big Hilbert space which
includes the zero-mode of $\xi$ and should be thought of as $(\xi +
\tilde{\xi}) \mathcal{O}$ where $\mathcal{O}$ is in the small Hilbert
space.

We will see in an example that this quantity appears to compute the change in
the closed string one-point function, just as is it does in the bosonic case.
However, because of the complexity of perturbation theory in the Berkovits superstring,
we do not have a general derivation of this result.

The organization of this paper is as follows: In section \ref{Review},
we review the construction of the invariants $W(\Psi,\mathcal{V})$, the
$\arctan(z)$ coordinate system and the closed string tadpole.  In
section \ref{ExplicitComputations}, we compute $W(\Psi,\mathcal{V})$
for marginal deformations and the tachyon vacuum.  In section
\ref{Derivation}, we show how the relation between the closed string
one-point function and $W(\Psi,\mathcal{V})$ can be derived from BRST
invariance of the closed string two-point function.  Finally, in
section \ref{susyversion} we discuss an extension to the
Berkovits superstring field theory.

\section{Review} \label{Review}

In this section we review the invariants $W(\Psi,\mathcal{V})$ introduced in
\cite{Hashimoto:2001sm,Gaiotto:2001ji} and discuss how they are computed in the
$\arctan(z)$ coordinates.  We then discuss some aspects of the closed
string tadpole diagram which will be useful later.

\subsection{The  invariants $W(\Psi,\mathcal{V})$}

\begin{figure}
\centerline{
\begin{picture}(380,122)(0,-10)
\includegraphics{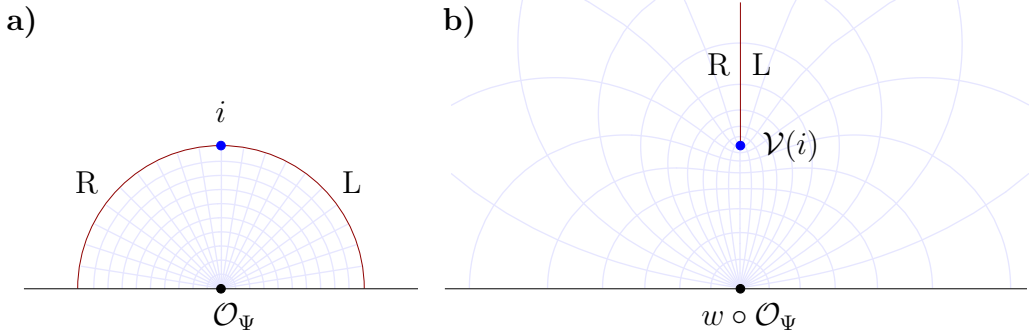}  
\end{picture}
\begin{picture}(0,0)(380,-10)
\put(68,-12){$\mathcal{O}_{\Psi}$}
\put(249,-12){ $w\circ \mathcal{O}_{\Psi}$}
\put(272,53){ $\mathcal{V}(i)$}
\put(69,65){$i$}
\put(117,38){L}
\put(16,38){R}
\put(272,83){L}
\put(255,83){R}
\put(-10,100){\bf a)}
\put(155,100){\bf b)}
\end{picture}
}
\caption{The construction of the invariants  $W(\Psi,\mathcal{V})$ is
shown. In a), we begin with a state $\Psi$ formed by inserting the
vertex operator $\mathcal{O}_{\Psi}$ into the UHP at the origin. The
wavefunction for the state $\Psi$ is to be thought of as living on the
unit semi-circle.  The left and right halves of the string as seen
from infinity are labeled L and R. The string midpoint is at $z = i$.
To contract the state with the identity, glue the semicircles L and R
together and map the resulting geometry to the plane using $z \to
w(z)$ as shown in b).  To saturate the ghostnumber, insert a closed
string field $\mathcal{V}$ at the midpoint, $w(i) =
i$. \label{MapToIdentity}}
\end{figure}

Consider a string field $\Psi$, defined as the state $|\Psi\rangle =
\mathcal{O}_\Psi(0)|0\rangle$, where $\mathcal{O}_\Psi$ is a ghost number 1
boundary operator and $|0\rangle$ is the $SL_2(\mathbb{R})$ vacuum. 
In the upper half plane, we may think of the state $|\Psi\rangle$ as living
on the unit semi-circle as in figure \ref{MapToIdentity}a.

To define the  invariants $W(\Psi,\mathcal{V})$, first map the upper half disk to the
entire upper half plane using the map,
\begin{equation}
  w(z) = \frac{2 z}{1-z^2} \ .
\end{equation}
This is shown in figure \ref{MapToIdentity}b. Next, to saturate the
ghost number on the UHP, add a ghostnumber 2 vertex operator
$\mathcal{V}(i)$ at the midpoint.  Finally, compute the correlator,
\begin{equation}
  W(\Psi,\mathcal{V}) = \langle \mathcal{V}(i) \, w\circ
  \mathcal{O}_\Psi \rangle_{\text{UHP}} \ .
\end{equation}
The key property of $W(\Psi,\mathcal{V})$ is that if $\mathcal{V}$ is
a weight $(0,0)$ primary, satisfying $\{Q_B, \mathcal{V}\} = 0$, then $W$
is invariant under the open string field theory gauge group,
\begin{equation}
   W(\Psi + Q_B \Lambda + [\Psi,\Lambda],\mathcal{V}) = W(\Psi,\mathcal{V}) \ .
\end{equation}
Since $W(\Psi,\mathcal{V})$ is linear in $\Psi$, this follows from the identities,
\begin{align}
   W(Q_B \Lambda,\mathcal{V}) &=0 \label{I1}\ , 
\\ W([\Psi,\Lambda],\mathcal{V}) &=0 \label{I2} \ .
\end{align}
To show (\ref{I1}), suppose $|\Lambda\rangle =
\mathcal{O}_\Lambda(0) |0\rangle$.  Then,
\begin{equation}
   W(Q_B \Lambda,\mathcal{V}) = \langle \mathcal{V}(i) \, w\circ
  \{Q_B,\mathcal{O}_\Lambda \} \rangle_{\text{UHP}}
 = -\langle [Q_B, \mathcal{V}(i)] \, w\circ
  \mathcal{O}_\Lambda \rangle_{\text{UHP}} = 0 \ ,
\end{equation}
where the second equality uses the BRST invariance of the boundary
conditions on the UHP, $\langle \{Q_B, \ldots \} \rangle = 0$.

The second identity (\ref{I2}) follows from essentially the same
arguments that show
\begin{equation}
   \int \Psi_1*\Psi_2 = \int \Psi_2*\Psi_1 \ .
\end{equation}
Assuming that $\mathcal{V}$ is a weight (0,0)
primary,
\begin{align}
   W(\Psi*\Lambda,\mathcal{V}) = \langle \mathcal{V}(i)
   \mathcal{O}_{\Psi*\Lambda} \rangle_{\text{UHP}} =\langle \mathcal{V}(i)\, f_1
   \circ \mathcal{O}_\Psi \, f_2 \circ \mathcal{O}_\Lambda \rangle_{\text{UHP}} \ ,
\label{PL}
\\
   W(\Lambda*\Psi,\mathcal{V}) = \langle \mathcal{V}(i)
   \mathcal{O}_{\Lambda*\Psi} \rangle_{\text{UHP}} =\langle \mathcal{V}(i)\, f_1
   \circ \mathcal{O}_\Lambda \, f_2 \circ \mathcal{O}_\Psi \rangle_{\text{UHP}} \ ,
\label{LP}
\end{align}
where
\begin{equation}
   f_1(z) = \frac{1+z}{1-z} \ , \qquad f_2(z) = -\frac{1-z}{1+z} \ .
\end{equation}
Noting that $f_1 = I\circ f_2$, where $I(z) = -1/z$ is the BPZ dual,
it follows that (\ref{PL}) and (\ref{LP}) are related by an $SL_2(\mathbb{Z})$
transformation and, hence, equal.  This implies
\begin{equation}
  W(\Psi*\Lambda - \Lambda*\Psi,\mathcal{V}) = 0 \ .
\end{equation}

\subsection{The $\arctan(z)$ frame}

It will be useful in the discussion to follow to know how to compute $W(\Psi,\mathcal{V})$ when 
the state $\Psi$ is given in the $\arctan(z)$ coordinate system that has played a prominent
role in recent developments.  Define,
\begin{equation}\label{fDef}
  \tilde z = f(z) = \tfrac{2}{\pi} \arctan(z) \ ,
\end{equation}
which takes the upper half plane to a semi-infinite cylinder of
circumference $2$.  A correlator on a semi-infinite cylinder of
circumference $n$ is defined by first rescaling $\tilde z \to \frac{2}{n} \tilde z$
to get back to a cylinder of width $2$ and then mapping $\tilde z \to
f^{-1}(\tilde z)$ to get back to the upper half plane.  We will often
follow the notation of \cite{Kiermaier:2007ba} and consider the fundamental region of
the cylinder to be the region $-\frac{1}{2} < \Re(\tilde z) < n-\frac{1}{2}$.
This unusual choice happens to be convenient for the form
of some string field solutions.

\begin{figure}
\centerline{
\begin{picture}(416,148)(0,-10)
\includegraphics{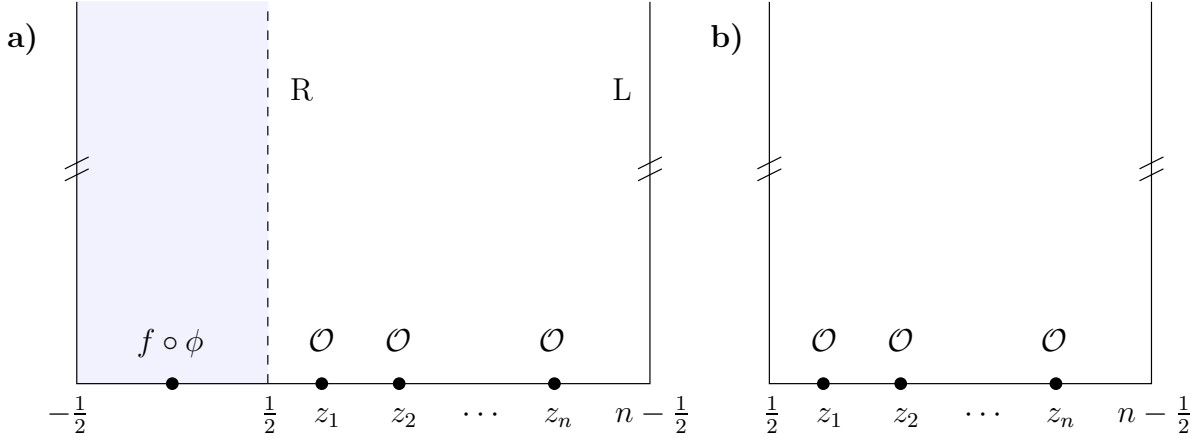}  
\end{picture}
\begin{picture}(0,0)(416,-10)
\put(24,15){$f\circ \phi$}
\put(-10,-12){$-\frac{1}{2}$}
\put(71,-12){$\frac{1}{2}$}
\put(91,-12){$z_1$}
\put(120,-12){$z_2$}
\put(147,-12){$\cdots$}
\put(178,-12){$z_n$}
\put(89,15){$\mathcal{O}$}
\put(118,15){$\mathcal{O}$}
\put(176,15){$\mathcal{O}$}
\put(205,-12){$n-\frac{1}{2}$}
\put(261,-12){$\frac{1}{2}$}
\put(281,-12){$z_1$}
\put(310,-12){$z_2$}
\put(337,-12){$\cdots$}
\put(368,-12){$z_n$}
\put(279,15){$\mathcal{O}$}
\put(308,15){$\mathcal{O}$}
\put(366,15){$\mathcal{O}$}
\put(395,-12){$n-\frac{1}{2}$}
\put(-25,130){\bf a)}
\put(241,130){\bf b)}
\put(82,110){R}
\put(204,110){L}
\end{picture}
}
\caption{In a) a typical state $|\Sigma\rangle$ is shown in cylinder
coordinates.  The shaded region represents the coordinate patch, or,
in other words, the image of the unit half disk under $f(z)$.  The
left and right halves of the state $|\Sigma\rangle$ are labeled L and
R. In b) $W(\Sigma,\mathcal{V})$ is shown.  This is obtained by
removing the coordinate patch and gluing the lines labeled by L and R
together.  As a final step the operator $\mathcal{V}$ should be
inserted at $i\infty$. \label{WOfCylindarState}}
\end{figure}

A prototypical state $|\Sigma\rangle$ defined in cylinder coordinates is
shown pictorially in \ref{WOfCylindarState}a.  Algebraically, we
define $|\Sigma\rangle$ through its overlap with an arbitrary test
state $\langle \phi|$,
\begin{equation}
  \langle \phi | \Sigma \rangle
 = \langle f \circ \phi(0) \, \,\mathcal{O}(\tilde z_1) \ldots \mathcal{O}(\tilde z_n) \rangle_{C_n}
\end{equation}
where the $\mathcal{O}$'s
are some local operators and the subscript $C_n$ indicates that the
correlator is to be evaluated on a cylinder of circumference $n$.  In
order for this to be a non-singular definition, we must require that
none of the $\tilde z_i$ are contained in the image of the unit
half-disk under the map $f(z)$.  This region is given by $-\frac{1}{2} \le
\Re(\tilde z) \le \frac{1}{2}$ (and its images under $\tilde z \to
\tilde z + n$).

Given a state $|\Sigma\rangle$ defined in this way, we would like to
compute $W(\Sigma,\mathcal{V})$.  The first step is to glue the left
and right halves of $\Sigma$ together.  In the $\tilde z$ coordinates,
the left and right halves of the string live at $\Re(\tilde z) = n
+\frac{1}{2}$ and $\Re(\tilde z) = \frac{1}{2}$ respectively as shown
in the figure.  To glue them together, we remove the coordinate patch
$-\frac{1}{2} < \Re(\tilde z)<\frac{1}{2}$, leaving us with a strip of
worldsheet of width $n-1$ and then glue the two sides of the
worldsheet together, giving us back a cylinder of circumference $n-1$.  This
is shown in figure \ref{WOfCylindarState}b.  Finally, the operator
$\mathcal{V}$ should be inserted at $i\infty,$ which is the string
midpoint in the $\tilde z$ coordinates.  In total, 
\begin{equation}
  W(\Sigma,\mathcal{V}) = 
\langle \mathcal{V}(i\infty) \, \mathcal{O}(\tilde z_1) \ldots \mathcal{O}(\tilde z_n)\rangle_{C_{n-1}} \ .
\end{equation}

\subsection{The closed string one-point function}

Since we wish to relate $W(\Psi,\mathcal{V})$ to the tree-level closed string
one-point function, it is useful to review how this diagram is
computed.  The closed-string one-point function is the amplitude with
one vertex operator $\mathcal{V}$ inserted on the disk.  Since there
are 3 CKVs on the disk, we may fix the position of the one vertex operator to the center of the disk, $z
= 0$.  Hence, $\mathcal{V}$ should be a fixed vertex operator of the form $c\tilde c
\mathcal{O}^{\text{m}}$ where $\mathcal{O}^{\text{matter}}$ is a
weight $(1,1)$ matter operator.  Note, however, that
\begin{equation}
  \langle \mathcal{V}(0) \rangle_{\text{disk}} = 0 \ ,
\end{equation}
since, to get a non-vanishing answer, we need soak up three ghost
zero-modes and we have only soaked up two.  The problem is that fixing
the position of $\mathcal{V}$ only removes two out of the three CKV's
and the third, which generates rotations of the disk, has an
associated ghost-zeromode.  Typically, if we have CKV's left over, a
diagram will vanish because the volume of the associated group of
symmetries is infinite.  In this case, the volume of the group of
rotations of the disk is just $2\pi$ so the amplitude is finite.

To soak up the remaining zero-mode, we add the ghost-measure corresponding to fixing one of the points $z = e^{i\theta}$ on the boundary of the disk.  Given an infinitesimal coordinate shift $\delta \sigma^a$,
its component along the boundary is given by
\begin{equation}
  \sin \theta \,\delta \sigma^1 -\cos \theta\,\delta \sigma^2  = -\Im(e^{-i\theta} \delta \sigma^z) \ ,
\end{equation}
at the point $z = e^{i \theta}$.  To get the correct measure, we
should then add\footnote{We are not attempting to determine the
overall sign of the ghost measure.  It has been picked to give
(\ref{MainRelation}) rather than
$\mathcal{A}_0^{\text{disk}}(\mathcal{V})
-\mathcal{A}_\Psi^{\text{disk}}(\mathcal{V}) $.}
\begin{equation}
  -\Im(e^{-i \theta} c(e^{i\theta})) = i e^{-i\theta} c(e^{i\theta})
\end{equation}
to ghost path integral.  The complete one-point function is given
by
\begin{equation}\label{OnePointFunction}
\mathcal{A}^{\text{disk}}(\mathcal{V}) = -\frac{e^{-i\theta}}{2\pi i} \langle \mathcal{V}(0) \,c(e^{i\theta})\rangle_{\text{disk}} \ .
\end{equation}
Note that we have included an extra factor of $(2\pi)^{-1}$ to account
for the volume of the CKV group.  One can check that
(\ref{OnePointFunction}) is independent of $\theta$ as it should be.
In general, we will pick $\theta = 0$.

\section{Computation of $W(\Psi,\mathcal{V})$ for known solutions}
\label{ExplicitComputations}

In this section, the invariants $W(\Psi,\mathcal{V})$ are computed for
various known solutions.  In each case, the result is found to be
consistent with the change in the one-point function of the closed
string under the shift from the original boundary conditions to the
new boundary conditions associated with the string field solution.

\subsection{Invariants of marginal deformations with trivial OPEs}
\label{TrivialOPE}

There are currently two (presumably) gauge-equivalent solutions to
the OSFT equations of motion that describe marginal deformations with
trivial OPE.  The first \cite{Schnabl:2007az,Kiermaier:2007ba}, which
is in Schnabl-gauge \cite{Schnabl:2005gv}, turns out to be impractical
for computing $W(\Psi,\mathcal{V})$.  The second state, discovered by Fuchs, Kroyter and Potting \cite{Fuchs:2007yy} and
Kiermaier and Okawa \cite{Kiermaier:2007vu}, appears to be more closely
related to the boundary  conformal field theory and is better suited for our
computation.  Their solution also has a natural extension to the
non-trivial OPE case, which we will take up in the next subsection.

The complete solution takes the form \cite{Kiermaier:2007vu},
\begin{equation}\label{KOsolution}
  \Psi^{\text{KO}} = \frac{1}{\sqrt{U}}(\Psi_L + Q_B) \sqrt{U} \ ,
\end{equation}
where $\Psi_L$ is a state to be introduced shortly and $U$ is a string
field whose form we will not need.  The state (\ref{KOsolution})
appears to be a gauge-transformation of the state $\Psi_L$; however,
neither $\Psi_L$ nor $U$ are real string fields so (\ref{KOsolution})
is not a proper gauge transformation.  Nevertheless, since,
$W(\Psi,\mathcal{V})$ has no knowledge of the reality condition, we
can work with the simpler state $\Psi_L$.

The state $\Psi_L$ is given by\footnote{In \cite{Kiermaier:2007vu}, this
would be written $\sum_{n = 1}^\infty \lambda^n \, \Psi^{(n)}_L$ as
they pick the opposite convention for the left and right halves of the
string wave function.  This affects the overall sign of the invariant as well as the sign of the deformation.},
\begin{equation}
  \Psi_L =- \sum_{n = 1}^\infty (-\lambda)^n \, \Psi^{(n)}_L \ ,
\end{equation}
where, following \cite{Kiermaier:2007vu}, we define the states $\Psi^{(n)}$ on a
cylinder of circumference $n+1$,
\begin{equation}
  \langle \phi | \Psi^{(n)}_L\rangle  = 
     \left\langle 
f\circ \phi(0) c J(1) \int_1^2 dt_1 \int_{t_1}^3 dt_2 \int_{t_2}^4 dt_3
\ldots
\int_{t_{n-2}}^n \, J(t_1) J(t_2) J(t_3) \ldots J(t_{n-1}) \right\rangle_{C_{n+1}} \ .
\end{equation}
As defined in (\ref{fDef}), the map $f$ is given by $f(z) = \frac{2}{\pi} \arctan(z)$.  The field
$J$ is assumed to be a weight $1$ primary boundary matter operator
with trivial OPE: $J(z) J(0) \sim \mathcal{O}(1)$.

To compute $W(\Psi^{(n)},\mathcal{V})$, remove the coordinate patch
$-1/2 < \Re(\tilde z) <1/2$ and re-glue to form a cylinder of width $n$.  Then insert $\mathcal{V}(i\infty)$:
\begin{multline}
 W(\Psi^{(n)},\mathcal{V}) =
 \\  
     \left\langle 
\mathcal{V}(i\infty) c J(0) \int_0^1 dt_1 \int_{t_1}^2 dt_2 \int_{t_2}^3 dt_3
\ldots
\int_{t_{n-2}}^{n-1} dt_{n-1}\, J(t_1) J(t_2) J(t_3) \ldots J(t_{n-1}) \right\rangle_{C_{n}} \ .
\end{multline}
Mapping this geometry to the disk using 
\begin{equation}
  g(\tilde z) = e^{2\pi i \tilde z/n} \ ,
\end{equation}
yields\footnote{Note that under $z \to \chi(z)$, a weight $h$ boundary
operator transforms as $\mathcal{O}(z) \to |\frac{\partial\chi}{\partial z}|^h
\mathcal{O}(\chi(z))$.}
\begin{equation}\label{KOintegral}
-i\left\langle \mathcal{V}(0) c J(1)
     \int_0^{\omega} dt_1 \int_{t_1}^{2\omega} dt_2
      \ldots \int_{t_{n-2}}^{(n-1)\omega}dt_{n-1} \, J(e^{i t_1})
     J(e^{i t_2})  \ldots J(e^{i t_{n-1}}) \right\rangle_{\text{disk}} \ ,
\end{equation}
where $\omega = 2\pi/n$.

Remarkably, as we will now demonstrate,  this complicated integral is equal to the simpler integral,

\begin{equation} \label{freeintegral}
-\frac{i}{2\pi n!}\left\langle \mathcal{V}(0)\, c(1)
     \int_0^{2\pi} dt_1 \int_{0}^{2\pi} dt_2
      \ldots \int_{0}^{2\pi} dt_{n}\, J(e^{i t_1})
     J(e^{i t_2})  \ldots J(e^{i t_{n}}) \right\rangle^{\text{m}}_{\text{disk}} \ .
\end{equation}
Notice that the difference between the two integration regions is that
in (\ref{KOintegral}) we have the constraints that $t_k \le \omega k$.
These inequalities are explained by the following
lemma:

\noindent
{\bf Lemma}: Given $n$ points on the unit circle, we may always label
them in counter clockwise order, $z_i = e^{i\theta_i}$, $i\in \{1,\ldots,n\}$ with increasing
$\theta_i$ such that
\begin{equation} \label{ineq}
  \theta_j-\theta_1 \le \tfrac{2\pi}{n} (j-1) \ .
\end{equation}

\noindent 
{\em Proof}: We use proof by contradiction.  Begin by extending the
definition of $\theta_i$ to include $i\in \mathbb{Z}$, by defining
$\theta_{i+n} = \theta_i + 2\pi$.  Assuming the lemma is false, we have that
for every $\theta_i$, there exists a $\theta_j$ with $j>i$ such that
\begin{equation}
  \theta_j - \theta_i > \tfrac{2\pi}{n} (j-i) \ .
\end{equation}
Hence, there exists a sequence $\{ \theta_{i_m}\}$ such that
\begin{equation}
   \theta_{i_{m}} - \theta_{i_{m-1}} > \tfrac{2 \pi}{n} (i_m - i_{m-1}) \ ,
\end{equation}
from which it follows that
\begin{equation} \label{thetaineq}
  \theta_{i_m} - \theta_{i_p} > \tfrac{2\pi}{n}(i_m - i_p) \ .
\end{equation}
Since there are only a finite number of points on the circle, there
must be two points in the sequence such that $i_a - i_b = k n$ for
some $k\in \mathbb{Z}$.  Since these represent the same point on the
circle, we learn that
\begin{equation}
  \theta_{i_a} - \theta_{i_b} = 2\pi k \ ,
\end{equation}
which is in contradiction with (\ref{thetaineq}) for $m= a$ and $p = b$. $\square$

The choice of $z_1$ is generically unique.  If there are two
possible points which may be chosen as the first point, it follows
from (\ref{ineq}) that they must be separated by an integer multiple of $2\pi/n$.

Now, consider the integral (\ref{freeintegral}).  Ignoring special points
in the integration region (which are measure zero), we can divide the
integral up into $n$ integrals in which one of the $n$ points is
picked to be $z_1$ and the rest of the points satisfy (\ref{ineq}).
We can fix the order of the remaining points at the expense of
introducing a factor of $(n-1)!$ and we may fix $z_1 = 1$ by a
rotation if we multiply the integral by $2\pi$ (which cancels the
$2\pi$ in (\ref{freeintegral})).  Finally, all of these $n$ integrals
are identical giving a factor of $n$ which combines with the $(n-1)!$
to cancel the $n!$ in (\ref{freeintegral}) giving (\ref{KOintegral}).

Summing up the terms in $W(\Psi,\mathcal{V})$ using (\ref{freeintegral}) gives

\begin{equation} 
 W(\Psi,\mathcal{V}) = -\frac{1}{2\pi i}\left\langle \mathcal{V}(0) \,c(1)
     \left[\exp\left(-\int_0^{2\pi} dt \,\lambda J(e^{it})\right) -1\right]\right\rangle_{\text{disk}} \ ,
\end{equation}
which, using (\ref{OnePointFunction}), is equivalent to
\begin{equation} 
 W(\Psi,\mathcal{V}) = \mathcal{A}^{\text{disk}}_{\Psi}(\mathcal{V}) - \mathcal{A}^\text{disk}_{0}(\mathcal{V}) \ .
\end{equation}
As defined in the introduction,
$\mathcal{A}^{\text{disk}}_{\Psi}(\mathcal{V})$ is the one-point
function with boundary conditions deformed by $\lambda J$.

\subsection{Invariants of marginal deformations with non-trivial OPE}

The preceding argument can be extended to the case with non-trivial
OPE in the case when the OPE takes the form,
\begin{equation}\label{marginalOPE}
  J(z) J(w) \sim \frac{1}{(z-w)^2} \ .
\end{equation}
The main change to the previous discussion is that the operators $J$
must be renormalized.  There are, however, some subtleties which we
dwell on here that more general readers may not be interested in and
we encourage them to skip to the next subsection.

In the non-trivial OPE case, the solution is again given in \cite{Fuchs:2007yy,Kiermaier:2007vu}.
We follow the notation of Kiermaier-Okawa \cite{Kiermaier:2007vu}.  Before we can introduce their
state, we need to describe their renormalization scheme.  This requires 
a number of definitions which we now repeat:

Let the Green's function on the cylinder be
denoted
\begin{equation}
  G(y_1,y_2) = \langle J(y_1)J(y_2) \rangle \ ,
\end{equation}
and, following \cite{Kiermaier:2007vu}, define the normal ordered
product,
\begin{equation}\label{normalorder}
  :\prod_{i = 1}^n J(y_i): = e^{-\frac{1}{2} \int dx_1 \,dx_2\, G(x_1,x_2) \frac{\delta}{\delta J(x_1)} \frac{\delta}{\delta J(x_2)} }
  \prod_{i = 1}^n J(y_i) \ .
\end{equation}
The object $\int_a^b dy\, J(y)$ appears often enough that it is useful
to define \cite{Kiermaier:2007vu}
\begin{equation}
  J(a,b) \equiv \int_a^b dy\, J(y) \ .
\end{equation}
To write down the marginal solution, we need to specify two
renormalized operators
\begin{equation}
  \left[ e^{-\lambda \, J(a,b)} \right]_r \ , \qquad \left[ J(a) e^{-\lambda J(a,b)} \right]_r \ .
\end{equation}
To do this, we need the renormalized correlators
\cite{Kiermaier:2007vu},
\begin{align}
\label{V2ren}
  \langle J(a,b)^2 \rangle_r &\equiv 2 \lim_{\epsilon \to 0} \left(
  \int_a^{b-\epsilon} dy_1 \int_{t_1+\epsilon}^b dy_2\, G(y_1,y_2) - \frac{b-\epsilon-a}{\epsilon} - \log \epsilon
  \right) \ ,
  \\
  \langle J(a) J(a,b)\rangle_r &\equiv \lim_{\epsilon \to 0}\left( \int_{a+\epsilon}^b dy\, G(a,y) - \frac{1}{\epsilon}\right) \ .
\end{align}
The full renormalized operators are  given by
\cite{Kiermaier:2007vu}
\begin{align}
  \left[ e^{-\lambda J(a,b)} \right]_r &\equiv e^{\frac{1}{2} \lambda^2 \langle J(a,b)^2 \rangle_r}
   :e^{-\lambda J(a,b)}: \ ,
   \\
    \left[ J(a) e^{-\lambda J(a,b)} \right]_r &\equiv e^{\frac{1}{2} \lambda^2 \langle J(a,b)^2 \rangle_r}
   :(J(a) - \lambda \langle J(a) J(a,b) \rangle_r )e^{-\lambda J(a,b)}: \ .
\end{align}
Note that these can be rewritten as
\begin{align} \label{renorm}
  \left[ e^{\lambda J(a,b)} \right]_r  = \lim_{\epsilon \to 0}
  R_{\epsilon} \exp \left(-  \lambda^2 (\log \epsilon-1)+
    \int_a^b dy \, (-\lambda J(y) - \frac{1}{\epsilon} \lambda^2)
  \right) \ ,
  \\
   \left[ J(a) e^{\lambda J(a,b)} \right]_r  = \lim_{\epsilon \to 0}
  R_{\epsilon} (J(a) + \frac{1}{\epsilon} \lambda)\exp \left(-\lambda^2 (\log \epsilon-1)+
    \int_a^b dy \, (-\lambda J(y) - \frac{1}{\epsilon}\, \lambda^2)
  \right) \ ,
\end{align}
where the operator $R_\epsilon$ removes all terms in which two $J$'s
are within $\epsilon$ of each other.  A few comments may help clarify
these choices.  Essentially, we are renormalizing $-\lambda J \to -\lambda J -
\frac{1}{\epsilon} \lambda^2$.  
However, note the first term in the exponential, $\chi =
-\lambda^2(\log \epsilon -1)$, which comes from
$\log\epsilon$ and finite piece subtracted off in (\ref{V2ren}).

The $e^\chi$ prefactor is unexpected from the point of view of the
renormalization of the boundary operator $J$ since only the
counterterm $\frac{1}{\epsilon} \lambda^2$ is needed in boundary
perturbation theory \cite{Recknagel:1998ih}.  Fortunately, all
dependence on $\chi$ will drop out when the full solution is
assembled.  

We define the powers $J^{(n)}(a,b)$ through the expansions (absorbing, as
in \cite{Kiermaier:2007vu}, the factors of $n!$),
\begin{equation}
  [e^{-\lambda J(a,b)}]_r = \sum_{n = 0}^\infty (-\lambda)^n [J^{(n)} (a,b)]_r \ , \qquad
  [J(a) e^{-\lambda J(a,b)}]_r = \sum_{n = 0}^\infty (-\lambda)^n [J(a) J^{(n)} (a,b)]_r \ .
\end{equation}

Define the states\footnote{To compare with \cite{Kiermaier:2007vu},
note that $A_L = A_0+\tilde A_0$.}
\begin{equation}
  U_{\alpha} \equiv \sum_{n = 0}^\infty (-\lambda)^n U_{\alpha}^{(n)} \ , \qquad
  A_{\alpha} = \sum_{n = 1}^\infty (-\lambda)^n A_{\alpha}^{(n)} \ , \qquad \tilde A_{\alpha} = \sum_{n= 2}^\infty (-\lambda)^{n}\tilde A_{\alpha}^{(n)} \ ,
\end{equation}
where
\begin{align}
\langle \phi| U_\alpha^{(n)} \rangle&= \langle f \circ \phi(0) \, [J^{(n)}(1,n+\alpha)]_r \rangle_{C_{n+\alpha+1}} \ ,
\\
\langle \phi| A_\alpha^{(n)} \rangle&= \langle f \circ \phi(0) \, [cJ(1) J^{(n-1)}(1,n+\alpha)]_r \rangle_{C_{n+\alpha+1}} \ ,
\\
\langle \phi| \tilde A_\alpha^{(n)} \rangle&= \tfrac{1}{2} \langle f \circ \phi(0) \, \partial c \,[ J^{(n-2)}(1,n+\alpha)]_r \rangle_{C_{n+\alpha+1}}
\ .
\end{align}
The complete marginal solution is given by\footnote{As in the trivial
OPE case, this solution does not satisfy the reality condition.
However, the real solution is once again gauge equivalent if we allow
complex gauge transformations.}
\begin{equation}
  \Psi = -(A_{0}+\tilde A_{0}) U_0^{-1} \ .
\end{equation}
Conveniently, if one computes the contribution of $\tilde A_0
U_0^{-1}$ to $W(\Psi,\mathcal{V})$, it is proportional to the ghost
correlator,
\begin{equation}
  \langle c\tilde c (i) \partial c(0) \rangle_{\text{UHP}} = 0 \ .
\end{equation}
Hence, we can ignore $\tilde{A}$ in our discussion and we need only compute
\begin{equation}
  W(\Psi,\mathcal{V}) = W(-A_0 U^{-1}_0,\mathcal{V}) \ .
\end{equation}

We now want to show that $A_0 U^{-1}_0$ contains only subtractions of
inverse powers of $\epsilon$ and that the contribution from $\chi =
-\lambda^2(\log \epsilon -1)$ does not enter.  To do this, define a
new renormalization $[\,]'_{r}$ in which the $\log \epsilon$ and
finite piece in (\ref{V2ren}) are not subtracted,
\begin{align}
  \left[ e^{-\lambda J(a,b)} \right]'_r  &= 
  R_{\epsilon} \exp \left(
    \int_a^b dy \, (-\lambda J(y) - \frac{1}{\epsilon}\, \lambda^2)
  \right) \ ,
  \\
   \left[ J(a) e^{-\lambda J(a,b)} \right]'_r  &=
  R_{\epsilon} (J(a) + \frac{1}{\epsilon} \lambda)\exp \left( \int_a^b dy \, (-\lambda J(y) - \frac{1}{\epsilon}\, \lambda^2)
  \right) \ .
\end{align}
Note that we can no longer take $\epsilon \to 0$ since these operators are not finite in that limit.
  Next, define $U'_{\alpha}$ and $A'_{\alpha}$ to be the same as $U_{\alpha}$ and $A_{\alpha}$ except using $[\,]'_r $ instead of $[\,]_r$.  We can express one in terms of the other as follows:
\begin{equation}
  U = \sum_{n =0}^\infty \chi^{n} U_{2n}' \ ,\qquad A_0 = \sum_{n = 0}^{\infty} \chi^{n} A'_{2n} \ .
\end{equation}
We then have
\begin{multline}
  A_0 U^{-1}_0 = \sum_{n = 0}^\infty \chi^n A_{2n}' (\sum_{m = 0}^\infty \chi^n U_{2m}')^{-1}
   \\
   = \sum_{n = 0}^\infty\sum_{N = 0}^{\infty} (-1)^N\left( \prod_{i = 1}^N \sum_{k_i = 1}^\infty \right)
   \chi^{n+k_1+\ldots +k_N}
   A'_{2n} (U'_0)^{-1} \prod_{i = 1}^N U_{2k_i}'(U'_0)^{-1} \ .
\end{multline}
Using the identity \cite{Kiermaier:2007vu},
\begin{equation}
  A'_{\alpha} (U'_0)^{-1} U'_{\beta} = A'_{\alpha+\beta} \ ,
\end{equation}
We find
\begin{equation}
  \sum_{n = 0}^\infty\sum_{N = 0}^{\infty} (-1)^N\left( \prod_{i = 1}^N \sum_{k_i = 1}^\infty \right)
   \chi^{n+k_1+\ldots +k_N}
   A'_{2n+2k_1+\ldots 2k_N} (U'_0)^{-1} \ .
\end{equation}
Note that the coefficient of $\chi^K A'_{2K}$ is 
\begin{equation} \label{chicoeff}
  \sum_{n = 0}^\infty\sum_{N = 0}^{\infty} (-1)^N\left( \prod_{i = 1}^N \sum_{k_i = 1}^\infty \right)
   \delta_{K,n+k_1+\ldots +k_N} \ .
\end{equation}
Replacing the Kronicker delta with a Dirac delta-function, we can write this as
\begin{multline}
  \sum_{n = 0}^\infty\sum_{N = 0}^{\infty} (-1)^N\left( \prod_{i = 1}^N \sum_{k_i = 1}^\infty \right)
   \delta(K-(n+k_1+\ldots +k_N))
   \\
    = \int_{-\infty}^\infty dy 
    \sum_{n = 0}^\infty\sum_{N = 0}^{\infty} (-1)^N\left( \prod_{i = 1}^N \sum_{k_i = 1}^\infty \right)
   e^{i y(K-(n+k_1+\ldots +k_N))} \ .
\end{multline}
Performing the sums over $n$ and $k_i$ gives
\begin{equation}
\int_{-\infty}^\infty dy 
    \sum_{n = 0}^\infty\sum_{N = 0}^{\infty} (-1)^N e^{iy(1+K)}\left( \frac{1}{e^{iy} - 1}\right)^{N+1}
    = \int_{-\infty}^\infty dy\, e^{i y K} = \delta(K) \ ,
\end{equation}
from which we learn (dividing by $\delta(0)$ if you will), that (\ref{chicoeff}) is just $\delta_{K,0}$.
We have found that
\begin{equation}
  A_0 U_0^{-1} = A'_0 (U'_0)^{-1} \ ,
\end{equation}
so that all $\chi$ dependence has dropped out as promised.  Note that,
since the left hand side is finite, the right hand side must be
finite.  This useful fact, which can be verified at low orders, tells
us that no $\log \epsilon$ terms ever arise in the full form of
$\Psi$.  This also implies that as far as $\Psi$ is concerned, we can
use $[ \, ]'_r$, which is the expected renormalization of $J$. We can
now write
\begin{multline}
  \langle \phi|A_0' (U'_0)^{-1} \rangle
   \\
   = 
   \sum_{n= 1}^\infty (-\lambda)^n
   \left\langle
   f\circ \phi\, \int_1^2 dy_1 \int_{y_1}^2 dy_1 \, \ldots \int_{y_{n-2}}^n dy_{n-1}
   \left[ cJ(0) J(y_1) J(y_2)\ldots J(y_{n-1}) \right]'_r \right\rangle_{C_{n+1}} \ .
\end{multline}
Inserting this state into $W$, the argument proceeds in the same manner as in the trivial OPE case.  We find, simply
\begin{equation} 
 W(\Psi,\mathcal{V}) = -\frac{1}{2\pi i}\left\langle \mathcal{V}(0) \,c(1)
     \left[\exp\left(-\int_0^{2\pi} dt \,\lambda J(e^{it})\right) -1\right]'_r\right\rangle_{\text{disk}} \ ,
\end{equation}
which, using (\ref{OnePointFunction}), gives
\begin{equation} 
 W(\Psi,\mathcal{V}) = \mathcal{A}^{\text{disk}}_{\Psi}(\mathcal{V}) - \mathcal{A}^\text{disk}_{0}(\mathcal{V}) \ .
\end{equation}
The only new feature here is that the boundary deformation generated by
$J$ has been renormalized using the appropriate counter term as
discussed in \cite{Recknagel:1998ih}.

\subsection{Invariants of the tachyon vacuum}

We can also compute the invariants for the tachyon vacuum solution.  The tachyon vacuum state is given by \cite{Schnabl:2005gv}
\begin{equation}
  \lim_{N\to \infty} \left( \psi_N - \sum_{n = 0}^N \partial_n \psi_n\right) \ ,
\end{equation}
where
\begin{equation}
  \langle \phi | \psi_k \rangle = \left \langle [f\circ \phi](0)
 \,c(-1) \left(\int_{-i\infty}^{i\infty} \frac{d\tilde z}{2\pi i} \,
 b(\tilde z)\right) c(1)
  \right\rangle_{C_{n+2}} \ .
\end{equation}

The invariant $W(\psi_n,c \bar c \mathcal{O}^{\text{m}})$ is given by
\begin{equation}
  \left\langle c(i\infty)c(-i\infty) \mathcal{O}^\text{m}(i\infty) \,\, 
c(n/2) \left(\int_{-i\infty}^{i\infty} \frac{d\tilde z}{2\pi i}\, b(\tilde z)\right) c(-n/2) \right\rangle_{C_{n+1}} \ .
\end{equation}
Applying
\begin{equation}
  g(\tilde z) = \tan\left( \frac{\pi \tilde z}{n+1} \right) \ ,
\end{equation}
we get
\begin{equation}
  \frac{n+1}{\pi} \frac{1}{(1+x^2)^2}
   \left\langle 
   c(i)c(-i) \mathcal{O}^\text{m}(i) \,\, 
   c(x) 
    \left(
       \int_{-i\infty}^{i\infty} \frac{dz}{2\pi i}\,(1+z^2) b(z)
    \right) 
   c(-x) 
 \right\rangle_{\text{UHP}} \ ,
\end{equation}
where $x = \tan(\frac{\pi}{2} \frac{n}{n+1})$.
Evaluating the ghost correlator, this reduces to
\begin{equation}
 W(\psi_n,c \tilde c\mathcal{O}^\text{m})= \frac{2i}{\pi}\langle \mathcal{O}^\text{m}(i) \rangle^\text{m}_{\text{UHP}} \ .
\end{equation}
Remarkably, this is independent of $n$.  It follows that
\begin{equation}
  W (\Psi,\mathcal{O}^{\text{m}}) = \lim_{N\to \infty} W (\psi_N - \sum_n \partial_n\psi_n,\mathcal{O}^{\text{m}})  = \lim_{N\to \infty} W (\psi_N,\mathcal{O}^{\text{m}})  \ ,
\end{equation}
which we can write as
\begin{equation} \label{tachyonvacinvariant}
 \frac{2 i }{\pi} \langle \mathcal{O}^{\text{m}}(i) \rangle^{\text{m}}_{\text{UHP}} =  \frac{1}{\pi} \langle c\tilde c \mathcal{O}(i) c(0)\rangle_{\text{UHP}} = \frac{1}{2\pi i} \langle \mathcal{V}(0) c(1) \rangle_{\text{disk}} = -\mathcal{A}^{\text{disk}}_0(\mathcal{V}) \ .
\end{equation}
It might seem surprising that the terms $\partial_n \psi_n$ would make
no contribution.  The reason for this simplification is that the sum,
$-\sum \lambda^n \partial_n \psi_n$ is a pure gauge state for $\lambda
<1$.  Since $W$ is gauge invariant, it follows that $W(\partial_n
\psi_n,\mathcal{O}^{\text{m}})$ must vanish for every $n$.

The result (\ref{tachyonvacinvariant}) should be interpreted as
\begin{equation} 
  W (\Psi,\mathcal{O}^{\text{m}}) 
     = \mathcal{A}_{\Psi}^\text{disk}(\mathcal{V})- \mathcal{A}_{0}^\text{disk}(\mathcal{V})\ ,
\end{equation}
where $\mathcal{A}_{\Psi}^\text{disk}(\mathcal{V})=0$ since there is
no source for closed strings in the tachyon vacuum.

\section{Derivation of the invariants from BRST invariance}\label{Derivation}

Having seen in two examples that $W(\Psi,\mathcal{V})$ computes the
closed string tadpole, it is desirable to find a general derivation of
this result.

Naively, one should begin with the usual method for finding a string
field theory diagram for a given amplitude: Open string field theory
diagrams are given by picking a minimal metric on the worldsheet
subject to the condition that any non-contractible Jordan open curves
have length at least $\pi$ \cite{Zwiebach:1990az}.  For a disk with one closed string
insertion and no open string insertions, however, there are no
non-contractible curves and the minimal metric surface has zero size.
Furthermore, including a background string field, representing a
change in the disk boundary conditions, it is not clear how to find
the appropriate minimal metric.

Although this direct approach fails, one can still try to use an
argument from BRST invariance: Consider a disk with {\em two} closed
string insertions and take the limit as the two insertions become
close together.  In this limit, the diagram is conformally equivalent
to a diagram in which the two closed string insertions are connected
to the boundary of the disk by a long tube.  If we pick the momenta of
the two closed string insertions such that the intermediate closed
string state is on-shell, this long tube will lead to a divergence
when we integrate over its length.  Conveniently, this divergence
gives rise to a BRST anomaly\footnote{Note that the diagram is neither
divergent, nor anomalous for generic momenta \cite{Klebanov:1995ni,Hashimoto:1996bf}.  See
\cite{Callan:1987px,Polchinski:1987tu} for a general discussion of how tadpoles can arise as surface terms in moduli space.} which is proportional
to the closed string tadpole diagram.

\begin{figure}
\centerline{
\begin{picture}(413,109)   
\includegraphics{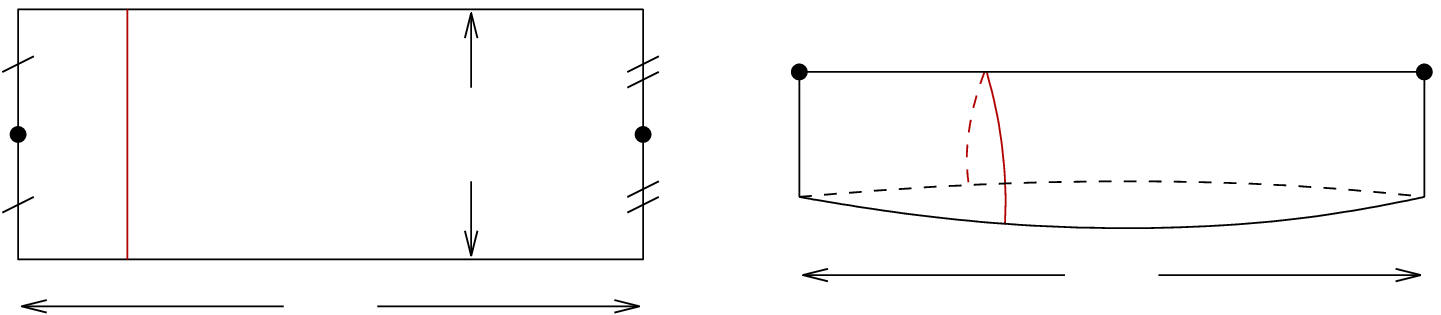}  
\end{picture}
\begin{picture}(0,0)(413,0)
\put(-18,49){$\mathcal{V}_1$}
\put(189,49){$\mathcal{V}_2$}
\put(210,72){$\mathcal{V}_1$}
\put(412,72){$\mathcal{V}_2$}
\put(87,-1){$T$}
\put(313,8){$T$}
\put(40,46){$b$}
\put(129,49){$\pi$}
\put(-30,95){\bf a)}
\put(200,95){\bf b)}
\end{picture}
}
\caption{ The closed string two-point function in the string field
theory conformal frame. There is one modulus, $T$, which is integrated
from $0$ to $\infty$. There is a single ghost insertion given by an
integral of the $b$-ghost over the red line. In a), the geometry is
shown as a flat strip with two identifications given by the hatches on
the right and left.  In b), the same geometry is shown in after the
identifications are performed.  Note the conical singularities at the
closed string insertions.  For consistency, $\mathcal{V}_{1,2}$ must
be weight $(0,0)$.\label{CSTwoPoint}}
\end{figure}

The closed string two-point function on the disk in the conformal
frame appropriate to string field theory is shown in figure
\ref{CSTwoPoint}
\cite{Shapiro:1987gq,Shapiro:1987ac,Freedman:1987fr,Zwiebach:1990az,Hashimoto:2001sm,Takahashi:2003kq,Garousi:2004pi}. The
amplitude is given by\footnote{Computations of the closed string
two-point function in open string field theory include
\cite{Takahashi:2003kq,Garousi:2004pi}.}
\begin{equation}\label{TwoPointFunction}
  \mathcal{A}(\mathcal{V}_1,\mathcal{V}_2) = 
  \langle \mathcal{I}|
   \mathcal{V}_1(i)\, 
     b_0 
    \int_{\epsilon/2}^{\infty} dT\,e^{-L_0 T}
    \mathcal{V}_2(i) 
  |\mathcal{I}\rangle \ ,
\end{equation}
where $\epsilon$ is a UV cutoff on the worldsheet, but an IR cutoff in
spacetime.  That this diagram is given by a propagator sandwiched
between two states will be very convenient when we repeat this
computation with a background open string field.

\begin{figure}
\centerline{
\begin{picture}(265,114)   
\includegraphics{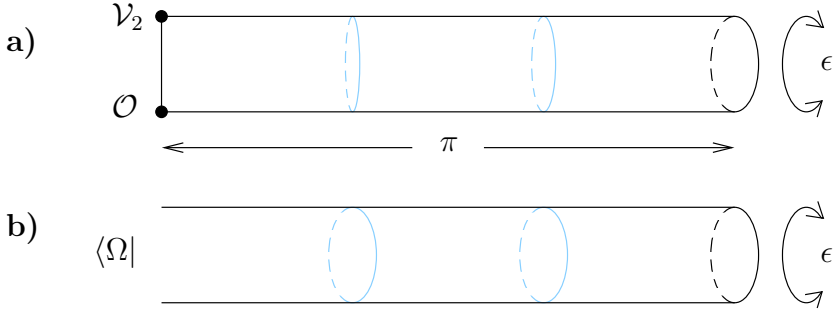}  
\end{picture}
\begin{picture}(0,0)(265,0)
\put(-20,73){$\mathcal{O}$}
\put(-20,108){$\mathcal{V}_2$}
\put(104,60){$\pi$}
\put(248,91){$\epsilon$}
\put(248,18){$\epsilon$}
\put(-27,19){$\langle \Omega |$}
\put(-60,98){\bf a)}
\put(-60,28){\bf b)}
\end{picture}
}
\caption{The surface term from replacing $\mathcal{V}_2 =
[Q_B,\mathcal{O}]$. In a), the amplitude (\ref{SurfaceTerm}) is
shown.  In b) the two closed string insertions are replaced with their
OPE. \label{OPEReplace}}
\end{figure}

To see the origin of the BRST anomaly, consider the case
when $\mathcal{V}_1 = Q_B \mathcal{O}$.  We then find,
\begin{multline}
  \langle \mathcal{I}|
   [Q_B,\mathcal{O}(i)]\, 
     b_0 
    \int_{\epsilon/2}^{\infty} dT\,e^{-L_0 T}
    \mathcal{V}_2(i) 
  |\mathcal{I}\rangle
\\
 = -\langle \mathcal{I}|
  \mathcal{O}(i)\, 
    \{Q_B, b_0 
    \int_{\epsilon/2}^{\infty} dT\,e^{-L_0 T} \}
    \mathcal{V}_2(i) 
  |\mathcal{I}\rangle
 = -\langle \mathcal{I}|
  \mathcal{O}(i)\, 
    \,e^{-L_0 T}\biggr|_{T = \epsilon/2}^\infty    \mathcal{V}_2(i) 
  |\mathcal{I}\rangle \ ,
\end{multline}
where we have used the properties $\{Q_B,b_0\} = L_0$ and the on-shell
condition $\{Q_B,\mathcal{V}_2\} = 0$ as well as $Q_B
|\mathcal{I}\rangle = 0$.  The contributions at $T \to \infty$ are not
relevant for the current discussion.  Dropping them gives
\begin{equation} \label{SurfaceTerm}
  \langle \mathcal{I}|
  \mathcal{O}(i)\, 
    \,e^{-L_0 \epsilon/2}   \mathcal{V}_2(i) 
  |\mathcal{I}\rangle \ .
\end{equation}
This amplitude is shown in figure \ref{OPEReplace}a.  Since $\epsilon$
is assumed to be very small, we may replace the two insertions of
$\mathcal{O}$ and $\mathcal{V}_2$ with their OPE, giving the geometry
in figure \ref{OPEReplace}b.  The geometry is considerably simplified.
We now have a closed string state, $|\Omega \rangle$ coming in
from in infinity and ending on a boundary.  Note that the OPE could
have singular terms since we are in a theory with tachyons. Such terms
correspond to propagation of the tachyon over long distances and
should be removed either by analytic continuation or explicit
subtraction.  In the absence of singularities, it follows that
$(L_0+\tilde{L}_0) |\Omega \rangle = 0$.  Note that if the OPE contains
no finite piece, the surface term vanishes.  This is why 
$\mathcal{O}$ and $\mathcal{V}_2$ must be tuned so that the intermediate
closed string state is on-shell.

Since $\Omega$ is overlapped with the $L_0+\tilde{L}_0 = 0$ part of
the boundary state, which is in the cohomology of $Q_B$, we may drop
the parts of $\Omega$ which are not physical; Hence, we may take\footnote{We are assuming that it is possible to divide the closed string fock space into two orthogonal pieces $\mathcal{H}_{\text{CFT}} = \mathcal{H}_{\text{coh}} \oplus \mathcal{H}_{\text{rest}}$ with the weight zero piece of the boundary state in the $Q_B$-cohomology, $\mathcal{H}_{\text{coh}}$.  Note that we have not shown that an {\em arbitrary} element of $\mathcal{H}_{\text{coh}}$ can be created from the OPE of the states $\mathcal{O}$ and $\mathcal{V}_2$, which would be required for a complete derivation.}
$\{Q_B,\Omega \} = 0$.  This is the closed string one-point
function which we wished to compute.  The point of this exercise is
that when we turn on an open string vev, we can repeat the same
computation to find the one-point function in the presence of an open
string field background.

\begin{figure}
\centerline{
\begin{picture}(448,148)   
\includegraphics{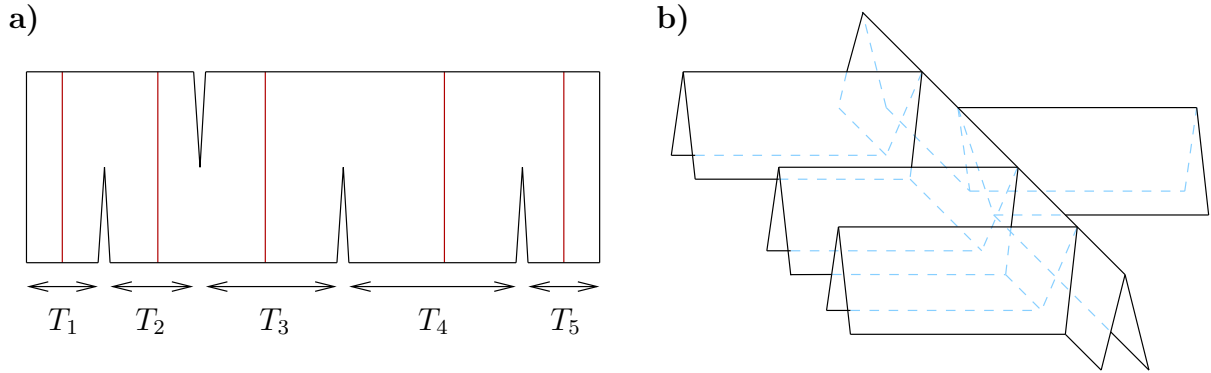}  
\end{picture}
\begin{picture}(0,0)(448,0)
\put(-10,130){\bf a)}
\put(235,130){\bf b)}
\put(5,16){$T_1$}
\put(38,16){$T_2$}
\put(85,16){$T_3$}
\put(145,16){$T_4$}
\put(195,16){$T_5$}
\end{picture}
}
\caption{The propagator in the presence of a open string field vev shown with four insertions of $\Psi_{\text{cl}}$. In
a), the insertions of $\Psi_{\text{cl}}$ are represented by cuts in
the worldsheet.  As shown in b), to get the full worldsheet geometry,
one must glue an infinitely long strip into each cut.
Each insertion of $\Psi_{\text{cl}}$ introduces one extra moduli in
addition to the modulus of the overall length of the propagator.  With
each modulus, one must add an integral of $b$ -- as shown in red in a) --
in order to get the right measure on moduli space.
\label{Propagator}}
\end{figure}

When we shift the vacuum $\Psi \to \Psi + \Psi_{cl}$, the only change
in the open string field theory action is a shift in the BRST operator,
\begin{equation}
  Q_B \to Q_B + [\Psi_{\text{cl}},\quad] \ .
\end{equation}
This introduces a term,
\begin{equation}
  \int \Psi * \Psi*\Psi_{\text{cl}} \ ,
\end{equation}
in the action which shifts the propagator.  The new propagator is given by
summing over all the ways of inserting $\Psi_{\text{cl}}$ into the old
propagator together with the appropriate ghost insertions.  This is illustrated in figure \ref{Propagator}.

Algebraically, the propagator between states $|A\rangle$ and $|B\rangle$
can be written as follows.  Define the adjoint action of $\Psi_{\text{cl}}$ by
\begin{equation}
  \ad_{\Psi_{\text{cl}}} \Phi = \Psi_{\text{cl}}*\Phi - (-1)^{\text{gh}(\Phi)} \Phi*\Psi_{\text{cl}} \ ,
\end{equation}
and
\begin{equation}
  D =  \int_0^\infty dT \, e^{-T L_0} \ .
\end{equation}
Then the full propagator is given by
\begin{equation}
  \sum_{n = 0}^{\infty} 
    \langle A| b_0 D \left(\ad_{\Psi_{\text{cl}}}\, b_0 D\right)^n |B\rangle \ .
\end{equation}

Given the propagator in the presence of $\Psi_{\text{cl}}$ one can
compute the modified closed-string two point function by replacing
the old propagator in (\ref{TwoPointFunction}) with the new one,
\begin{equation} \label{TwoPointFull}
 \mathcal{A}_\Psi(\mathcal{V}_1,\mathcal{V}_2) =  
    \sum_{n = 0}^\infty\langle \mathcal{I}|\, \mathcal{V}_1(i) 
      \,\, b_0 D \left( \ad_{\Psi_{\text{cl}}} b_0 D \right)^n \mathcal{V}_2 (i)|\mathcal{I}\rangle
\end{equation}
To extract the one-point function, again replace $\mathcal{V}_1 =
\{Q_B,\mathcal{O}\}$.  After some algebra and using the equations of motion for
$\Psi_{\text{cl}}$ one finds (see appendix \ref{SurfaceTermComp} for details):
\begin{equation} \label{TwoPointFullSurface}
-\int_{\epsilon/2}^\infty dT\,\frac{\partial}{\partial T}
 \sum_{n = 0}^\infty\left( \prod_{i = 0}^n \int_0^\infty dT_{n}\right)\, \delta (T-\sum_{i = 0}^n T_i)
 \langle \mathcal{I}|\, \mathcal{O}(i) 
      \,\,  D_{T_0} \left( \prod_{i = 1}^n \{b_0, \ad_{\Psi_{\text{cl}}}\}  D_{T_i} \right) \mathcal{V}_2(i)|\mathcal{I}\rangle \ ,
\end{equation}
where 
\begin{figure}
\centerline{
\begin{picture}(426,206)(0,-10)   
\includegraphics{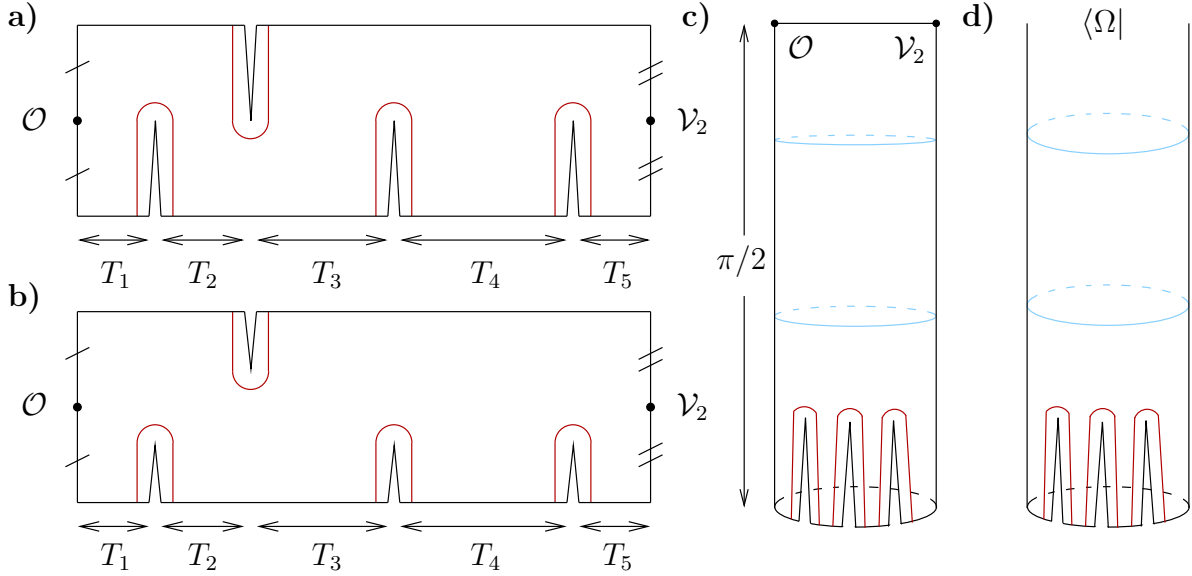}  
\end{picture}
\begin{picture}(0,0)(426,-10)
\put(-20,153){$\mathcal{O}$}
\put(-20,45){$\mathcal{O}$}
\put(228,153){$\mathcal{V}_2$}
\put(228,45){$\mathcal{V}_2$}
\put(10,95){$T_1$}
\put(43,95){$T_2$}
\put(90,95){$T_3$}
\put(150,95){$T_4$}
\put(200,95){$T_5$}
\put(10,-12){$T_1$}
\put(43,-12){$T_2$}
\put(90,-12){$T_3$}
\put(150,-12){$T_4$}
\put(200,-12){$T_5$}
\put(270,180){$\mathcal{O}$}
\put(310,180){$\mathcal{V}_2$}
\put(381,190){$\langle \Omega |$}
\put(243,100){$\pi/2$}
\put(-25,192){\bf a)}
\put(-25,86){\bf b)}
\put(230,192){\bf c)}
\put(336,192){\bf d)}
\end{picture}
}
\caption{Various representations of the surface term are shown for the
case of four insertions.  In a), a representation of
(\ref{surfaceTerm}) is given. It is assumed that $\sum T_i =
\epsilon/2$. In this form the $\epsilon\to 0$ limit is difficult because
the operators $\mathcal{O}$ and $\mathcal{V}_2$ collide with the ends
of the cuts and the $b$-ghost insertions.  In b) a reparametrization for
the classical solution $\Psi_{\text{cl}}$ is used so that the cuts
do no reach the midpoint of the string. Performing the identifications
in b) produces the diagram c) which now has a long tube separating the
operators $\mathcal{O}$ and $\mathcal{V}_2$ from the cuts.  As shown
in d), when $\epsilon$ is small we can replace the top of the diagram with a
single closed string state, $|\Omega\rangle$.
\label{SurfaceTermWithPsi}}
\end{figure}
\begin{equation}
  D_{T_i} = b_0 e^{-T_i L_0} \ .
\end{equation}
This leads to the surface term,
\begin{equation} \label{surfaceTerm}
 \sum_{n = 0}^\infty\left( \prod_{i = 0}^n \int_0^\infty dT_{n}\right)\, \delta (\epsilon/2-\sum_{i = 0}^n T_i)
 \langle \mathcal{I}|\, \mathcal{O}(i) 
      \,\,  D_{T_0} \left( \prod_{i = 1}^n \{b_0, \ad_{\Psi_{\text{cl}}}\}  D_{T_i} \right) \mathcal{V}_2(i)|\mathcal{I}\rangle \ .
\end{equation}
Geometrically, this amplitude is given by figure
\ref{SurfaceTermWithPsi}a.  It is important to point out that that the
cutoff $\epsilon$ is not conformally/BRST invariant so the expression
(\ref{surfaceTerm}) is not invariant under gauge transformations of
$\Psi_{\text{cl}}$ except in the limit $\epsilon \to 0$.

Unlike in the case without an open string background, it is not clear that, when $\epsilon$ is very small in
(\ref{surfaceTerm}), one can replace $\mathcal{O}$ and
$\mathcal{V}_2$ with their OPE.  The problem is that the two closed
string operators are not separated from the rest of the geometry by a
long tube.  Instead, the midpoints of the $\Psi_{\text{cl}}$ insertions and
integrals of the $b$-ghost all remain close to the closed string
insertions.

To fix this problem, one can perform a gauge transformation of
$\Psi_{\text{cl}}$ which reduces its height.  This reparametrizaion,
which is discussed in appendix \ref{reparamdiscussion}, allows one to
make a cut in the propagator which is some height $h<\pi/2$ and insert strip representing $\Psi_{\text{cl}}$ which has
been shrunk by a factor of $2h/\pi$.  Since, as mentioned above, the
amplitude is not invariant under gauge transformations for finite
$\epsilon$, this step may seem suspicious.  However, as will be seen
in a moment, gauge invariance will be restored in the small $\epsilon$
limit and the dependence on $h$ will drop out.

The amplitude with the gauge transformed $\Psi_{\text{cl}}$'s is shown
in figure \ref{SurfaceTermWithPsi}b.  Performing the identifications
leads to a geometry shown in figure \ref{SurfaceTermWithPsi}c.  As can
be seen from the figure, there is now a long tube separating the
closed string insertions from the rest of the geometry so one may
replace them with their OPE as shown in figure
\ref{SurfaceTermWithPsi}d.  One can then check that, assuming we can
drop the non-physical parts of $\Omega$, so that $Q_B |\Omega \rangle
= 0$, the gauge invariance $\Psi_{\text{cl}}\to \Psi_{\text{cl}} + Q_B
\Lambda + [\Psi_{\text{cl}},\Lambda]$ is restored\footnote{It is nice
to have an independent check that this amplitude is the closed string
one-point function.  Here is a sketch of an alternate argument: since
gauge invariance is restored, we can reparametrize the width of the
state $\Psi_{\text{cl}}$ to limit it to an identity state with a
single operator $c \mathcal{O}$ inserted on the boundary.  Using the
$b$-integrals to remove the $c$ ghost, we are left with a disk with
the boundary deformation $\exp(\int \mathcal{O})$.  As one can check
in simple cases, this typically generates the renormalized boundary
deformation associated with the state $\Psi_{\text{cl}}$ so that the
diagram reduces to $\mathcal{A}_{\Psi}^{\text{disk}} (\mathcal{V})$.}.

By unitarity, the amplitude pictured in figure
\ref{SurfaceTermWithPsi}d should be the closed string one-point
function on a disk with boundary conditions CFT${}_{\Psi_{\text{cl}}}$.
We may suppose, without loss of generality, that 
\begin{equation}
  \Omega = (\partial c - \bar \partial \tilde c)c \tilde c \mathcal{O}^{\text{m}} \ ,
\end{equation}
where $\mathcal{O}^{\text{m}}$ is a weight $(1,1)$ primary.  Set
$c\tilde c \mathcal{O}^{\text{m}} = \mathcal{V}$.  The vertex operator
$\Omega$ is ghost number 3.  The extra ghostnumber corresponds to
fixing the CKV corresponding to the rotation of the cylinder.  To write
the amplitude in terms the standard ghostnumber 2 operator
$\mathcal{V}$, pull one of the $b$-ghost integrals off of the bottom
of the cylinder and push it up till it encircles the state $|\Omega\rangle$.
Next, let the $b$-ghost integral act on $|\Omega\rangle$ giving
\begin{equation}
  \frac{2\pi}{\epsilon }(b_0 - \tilde{b}_0) |\Omega \rangle = 
        \frac{2\pi}{\epsilon} |\mathcal{V}\rangle \ .
\end{equation}
The $\epsilon^{-1}$ can be used to fix the location of the cut
whose $b$-ghost integral we removed since, by rotational invariance,
the integral over its position just gives a factor of $\epsilon$.

At this point, the amplitude still bears little resemblance to the
 invariants $W(\Psi,\mathcal{V})$.  However, it turns out that by
simultaneously increasing the height $h$ of the insertions {\em and}
rescaling the wedge width on which the state $\Psi_{\text{cl}}$ is
defined, the amplitude dramatically simplifies.  To see why, consider
the state $\Psi_{cl}$ to be defined in the $\arctan(z)$ coordinates.
To map $\Psi_{cl}$ to the strip coordinates appropriate for gluing
$\Psi_{\text{cl}}$ to the cylinder, we should use
\begin{equation}
  \xi(z) =\frac{2h}{\pi}\log (\tan (\frac{\pi \tilde z}{2})),
\end{equation}
where the factor of $h$ accounts for the change in height of the insertion.  Suppose
that, in addition to changing the height of the solution, we also reparametrize it
by changing its {\em width}.  This can be accomplished by rescaling the state using $\tilde z\to \rho \tilde z$
while leaving the coordinate patch alone.  This is the standard reparametrization of the
wedge width discussed, for example in \cite{Schnabl:2002gg,Schnabl:2002ff,Rastelli:2006ap,Okawa:2006sn}.

\begin{figure}
\centerline{
\begin{picture}(353,122)(0,-10)   
\includegraphics{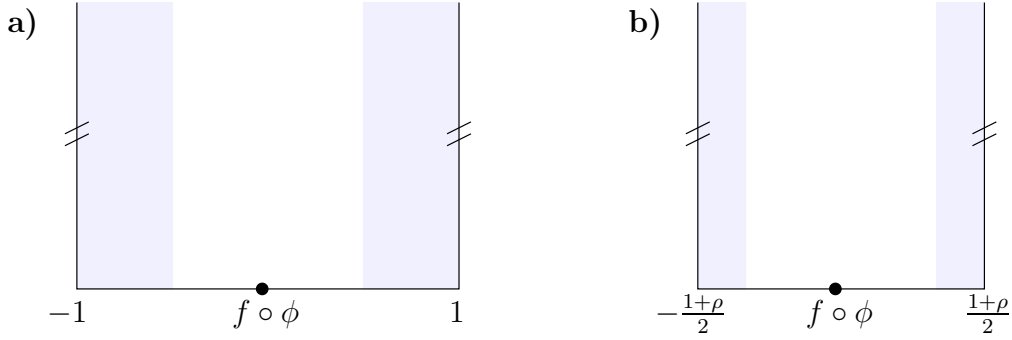}  
\end{picture}
\begin{picture}(0,0)(353,-10)
\put(60,-10){$f\circ \phi$}
\put(142,-10){$1$}
\put(-10,-10){$-1$}
\put(277,-10){$f\circ \phi$}
\put(337,-10){$\frac{1+\rho}{2}$}
\put(220,-10){$-\frac{1+\rho}{2}$}
\put(-25,100){\bf a)}
\put(210,100){\bf b)}
\end{picture}
}
\caption{Reparametrization of the wedge width.  In a), a standard
state is given in the $\arctan(z)$ coordinates.  In b), the state is
shrunk by a factor of $\rho$ while the coordinate patch is left alone
giving a cylinder of width $1+\rho$.
\label{widthReparam}}
\end{figure}

\begin{figure}
\centerline{
\begin{picture}(263,177)(-30,-10)   
\includegraphics{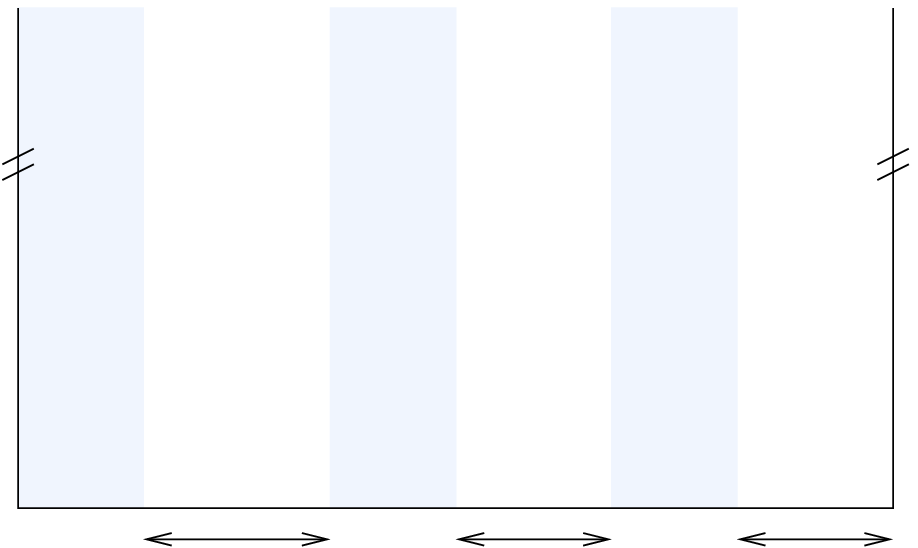}  
\end{picture}
\begin{picture}(0,0)(233,-10)
\put(12,20){$\Psi_{\text{cl}}$}
\put(60,-12){$T_1$}
\put(93,20){$B_1 \Psi_{\text{cl}}$}
\put(144,-12){$T_2$}
\put(175,20){$B_1 \Psi_{\text{cl}}$}
\put(227,-12){$T_3$}
\put(120,170){$|\mathcal{V}\rangle$}
\put(-120,86){\parbox{1in}{\begin{equation*}\int_0^\infty dT_i\,\,\delta(\sum_{i} T_i - \epsilon)\end{equation*}}}
\end{picture}
}
\caption{The resulting geometry for the case of three
$\Psi_{\text{cl}}$ insertions after flattening the insertions of
$\Psi_{\text{cl}}$ using a double gauge transformation.  The field
$\Psi_{\text{cl}}$ is now inserted into the geometry in the $\arctan$
coordinates.  The $b$-ghost integrals have become $B_1$'s acting on
all but one of the $\Psi_{\text{cl}}$'s.
\label{FlattenedGeometry}}
\end{figure}

In detail, suppose we take the original state to be defined on a
cylinder of circumference $2$ as shown in figure \ref{widthReparam}a.
Shrinking the wedge width by taking $\tilde z\to \rho \tilde z$ while leaving the
coordinate patch alone defines a new state $\Psi_{\text{cl}}'$ which
is shown in figure \ref{widthReparam}b.  The full map from the
original state $\Psi_{\text{cl}}$ to the coordinates we are using for gluing is
then given by
\begin{equation}
  \xi'(z) = \frac{2h}{\pi}\log \left[\tan \left(\tfrac{\pi}{2} ((\tilde z-\tfrac{1}{2}) \rho + \tfrac{1}{2})\right)\right] \ .
\end{equation}
The limit we are interested in is taking $h \to \infty$ with $\rho =
1/2h$. Focusing on the region of worldsheet near $\tilde z = 1/2$ (to avoid the branchcut of the $\log$), one can verify that
\begin{equation}
 \lim_{h\to \infty}  \frac{2h}{\pi}\log \left[\tan \left(\tfrac{\pi}{2} ((\tilde z-\tfrac{1}{2})\tfrac{1}{2h} + \tfrac{1}{2})\right)\right] = \tilde z-\frac{1}{2} \ ,
\end{equation}
which is just a simple translation.  In other words, in the limit $h
\to \infty$ $\rho \to 0$, $h\rho =1/2$, a state $\Psi_{\text{cl}}$ as defined
in the $\arctan(z)$ coordinates should be inserted into the cylinder
geometry by cutting a infinite vertical strip in the cylinder and
gluing in $\Psi_{\text{cl}}$ {\em without any conformal
transformations}.  The general picture is shown in figure \ref{FlattenedGeometry}\footnote{This representation of a string field theory amplitude is reminiscent of \cite{Fuji:2006me,Rastelli:2007gg,Kiermaier:2007jg}.}

In the resulting geometry, the integrals over the $b$-ghost just become
the operator $B_1 = b_{-1}+b_{1}$, which, in cylinder coordinates, is 
\begin{equation}
  \arctan \circ B_1 = \oint \frac{d\tilde z}{2\pi i} b(\tilde z) \ .
\end{equation}
The important point to note is that using the double gauge
transformation, we have flattened out the conical singularities that
arose from inserting $\Psi_{\text{cl}}$ into the cylinder geometry.
This allows one to act with $B_1$ on $\Psi_{\text{cl}}$ in the
obvious way.

One might worry about two problems in this limit: First, although the
curvature singularities are disappearing as we increase the height and
decrease wedge width, we are nonetheless bringing a curvature
singularity near the insertion of $\mathcal{V}$.  We believe that,
because $\mathcal{V}$ is a weight zero primary, there should be no
divergences from this limit.  Second, increasing the height of the
insertions pushes the contour integrals of $b$ close to $\mathcal{V}$.
Here again we believe there should be no singularity since the
$b$-integral contours can be made to go through $\mathcal{V}$ without
any divergence as can be checked by mapping the geometry to a disk.
(Note that this would not have been true before we removed a
$b$-integral from one of the $\Psi_{\text{cl}}$ insertions and let it
act on the closed string state). We fully admit, however that this
double reparametrizaion is delicate and additional operators inside the state
$\Psi_{\text{cl}}$ could also create potential divergences.

With these caveats in mind, consider taking the $\epsilon \to 0$
limit.  First, note that the worldsheet does not become singular
anywhere in this limit since the $\Psi_{\text{cl}}$ insertions can be
assumed to have a finite minimum thickness.  Furthermore, there are no
singularties when $\Psi_{\text{cl}}$ insertions become close as $B_1
\Psi_{\text{cl}} * \Psi_{\text{cl}}$ and $ \Psi_{\text{cl}} * B_1
\Psi_{\text{cl}}$ are finite\footnote{This is true at least for the
known solutions.  Since there is, at present, no general ``regularity
condition'' on the string field, we cannot say if this assumption is
always true, even if it seems reasonable.}.  However, the
integration regions go to zero size in this limit, so each term with
more than one $\Psi_{\text{cl}}$ will vanish.

The only terms that remain, are the case with one $\Psi_{\text{cl}}$ which
we recognize as the invariant $W(\Psi_{\text{cl}},\mathcal{V})$ and the case with no $\Psi_{\text{cl}}$'s
which is just the one-point function with $\Psi_{\text{cl}} =0 $.
Hence, we have found
\begin{equation}
  \mathcal{A}_{\Psi}^{\text{disk}}(\mathcal{V}) = \mathcal{A}_{0}^{\text{disk}}(\mathcal{V}) + W(\Psi,\mathcal{V}) \ ,
\end{equation}
which reproduces (\ref{MainRelation}).

  \section{Extension to Berkovits' open superstring field theory} \label{susyversion}

In this section, the extension to the Berkovits
superstring field theory of the  invariants $W(\Psi,\mathcal{V})$  is discussed.  The invariants are computed for
the case of marginal deformations with trivial OPE, yielding a
formula for the invariants in terms of the closed string one-point
function analogous to the bosonic case.

\subsection{A gauge-invariant observable for the superstring}

To extend to the superstring case, one needs an object which is
invariant under the modified gauge trasformation,
\begin{equation} \label{supergauge}
  e^{\Phi} \to e^{Q_B \Lambda} e^{\Phi} e^{\eta_0 \Lambda'} \ ,
\end{equation}
where $\Lambda$ and $\Lambda'$ are two gauge parameters.  Such an
invariant was written down in \cite{Michishita:2004rx}.  Here we take
a slightly different, but equivalent, approach\footnote{The invariant
written down in \cite{Michishita:2004rx} is simply $\langle
\mathcal{I}|\mathcal{V}(i) |\Phi\rangle$, with $Q_B \mathcal{V} =
\eta_0 \mathcal{V} = 0$.  Our invariant gives $\langle
\mathcal{I}|\mathcal{V}(i) |e^{-\Phi}Q_B e^{\Phi}\rangle = \langle
\mathcal{I}|\mathcal{V}(i) |Q_B \Phi\rangle = \langle
\mathcal{I}|\{Q_B,\mathcal{V}(i)\} |\Phi\rangle$, which, given our
assumptions on $\mathcal{V}$, reduces to the same thing.  The
advantage of our form comes from the fact that many superstring
solutions are found by guessing $\Omega$ and then later finding
$\Phi$, which is often much more complicated.}.

Define
\begin{equation} \label{omegadef}
  \Omega = e^{-\Phi} Q_B e^{\Phi}.
\end{equation}
The field $\Omega$ transforms under (\ref{supergauge}) as
\begin{equation} \label{omegatransformation}
  \Omega \to e^{-\eta_0\Lambda'} \left(\Omega +  Q_B\right) e^{\eta_0 \Lambda'} \ .
\end{equation}
Notice that it is invariant under the transformations generated by
$\Lambda$.  Consider the object,
\begin{equation}\label{susyinvariant}
  \widehat{W}(\Phi,\mathcal{V}) = \langle \mathcal{I} | \mathcal{V}(i)
  | \Omega(\Phi)\rangle \ ,
\end{equation}
where $\mathcal{V}$ is a weight $(0,0)$ primary. If $\mathcal{V}$
satisfied $Q_B \mathcal{V} = 0$ then we would find that $\widehat{W} =
0$ since by (\ref{omegadef}) $\Omega$ is pure-gauge in the bosonic sense.
We instead assume that
\begin{equation}
  Q_B (\eta_0+\tilde{\eta}_0) \mathcal{V} 
   = (\eta_0+\tilde{\eta}_0) Q_B \mathcal{V}= 0 \ , \qquad Q_B \mathcal{V} \ne 0  \ .
\end{equation}
We can now check that (\ref{susyinvariant}) is invariant under
(\ref{omegatransformation}). To see this, note that under the gauge
transformation (\ref{supergauge}),
\begin{equation} \label{Wshift}
  \widehat{W}(\Omega,\mathcal{V}) \to \widehat{W}(\Omega,\mathcal{V}) + 
\widehat{W}(e^{-\eta_0 \Lambda'} Q_B e^{\eta_0 \Lambda'},\mathcal{V}) 
\end{equation}
To show that the second term vanishes, define
\begin{equation}
  \Sigma_{\tau} = e^{-\tau \eta_0 \Lambda} Q_B e^{\tau \eta_0 \Lambda} \ ,
\end{equation}
and consider
\begin{multline}
\partial_\tau \langle \mathcal{I} |\mathcal{V}(i) |\Sigma_\tau\rangle
 = \langle \mathcal{I} |\mathcal{V}(i) | \left(Q_B \eta_0 \Lambda  
   + [\Sigma_\tau,\eta_0 \Lambda]\right)\rangle 
\\
= \langle \mathcal{I} |\mathcal{V}(i) |Q_B \eta_0 \Lambda  \rangle 
   = \langle \mathcal{I} |Q_B (\eta_0 + \tilde{\eta}_0)\mathcal{V}(i) 
        | \Lambda  \rangle = 0 \ .
\end{multline}
Since $\Sigma_0 = 0$, it follows that
\begin{equation}
  \langle \mathcal{I} |\mathcal{V}(i) |\Sigma_\tau\rangle = 0 \ .
\end{equation}
Since $\Sigma_1$ is the shift term in the gauge transformation
(\ref{Wshift}), $\widehat{W}(\Phi,\mathcal{V})$ is gauge
invariant under (\ref{supergauge}).

\subsection{Computation of $\widehat{W}$ for marginal solutions with trivial OPE}

For marginal solutions with trivial OPE there are two known solutions for the Berkovits superstring field theory.
The first, found by Erler and Okawa \cite{Erler:2007rh,Okawa:2007ri},
is similar to the Schnabl gauge solution in the bosonic theory and
does not appear to be simple to work with in this context.  The
second, found by Fuchs and Kroyter \cite{Fuchs:2007gw} and Kiermaier and Okawa \cite{Kiermaier:2007ki}, which is
analogous to their bosonic solutions, is, once again, more practical for our considerations.

Following the notation of Kiermaier and Okawa \cite{Kiermaier:2007ki}, let $\widehat{V}_{1/2}$ be a
superconformal primary with weight $1/2$ and define $\widehat{V}_1 =
G_{-1/2} \widehat{V}_{1/2}$. Putting
\begin{equation} \label{susyO}
  \mathcal{O}_L = c \widehat{V}_1 + \eta e^\phi \widehat{V}_{1/2} \ ,
\end{equation}
an exact solution for $\Psi_L = e^{-\Phi} Q_B e^{\Phi}$ can be written as
\begin{equation}
  \Psi_{L} = -\sum_{n = 1}^\infty (-\lambda)^n \Psi_{L}^{(n)} \ ,
\end{equation}
where
\begin{equation}
  \langle \phi |\Psi_L^{(n)}\rangle = \left\langle f\circ \mathcal{\phi}(0)
   \mathcal{O}_L(1) \prod_{m = 2}^n \int_{t_{m-1}}^m dt_m \,\widehat{V}_1(t_m) \right\rangle_{C_{n+1}} \ ,
\end{equation}
and $t_1 \equiv 1$.  One can now compute the invariant
$\widehat{W}(\Psi_L,\mathcal{V})$ in a similar fashion to the bosonic case.
For an NS-NS closed string field, we can represent $\mathcal{V}$ by
\begin{equation}
  \mathcal{V} = (\xi + \tilde \xi) c \tilde c e^{-\phi-\tilde \phi} \mathcal{O}^{(\frac{1}{2},\frac{1}{2})} \ ,
\end{equation}
where $\mathcal{O}^{(\frac{1}{2},\frac{1}{2})}$ is a weight
$(\frac{1}{2},\frac{1}{2})$ matter primary.  On the disk
\begin{equation}
  \widehat{W}(\Psi_L,\mathcal{V})
 = i\sum_{n = 1}^\infty (-\lambda)^n\left\langle
  \mathcal{V}(0) \mathcal{O}_L (1) \prod_{m = 2}^n 
\int_{\theta_{m-1}}^{2\pi \frac{m-1}{n}} d\theta \,\widehat{V}_1 (e^{i\theta_m})
\right \rangle_{\text{disk}} \ .
\end{equation}
Examining the $\xi \eta$
ghost system reveals that we can replace $\mathcal{O}_L$ with just its
first term $c \widehat{V}_1$ since the second term will make no
contribution.  The $\eta\xi$ part of the amplitude becomes simply
$\langle \xi(z) + \tilde \xi(\bar z)\rangle =2$, saturating the $\xi$ zeromode.
We thus find,
\begin{equation}
  \widehat{W}(\Psi_L,\mathcal{V})
 = i\sum_{n = 1}^\infty (-\lambda)^n\left\langle
  \mathcal{V}(0) c\widehat{V}_1 (1) \prod_{m = 2}^n \int_{\theta_{m-1}}^{2\pi \frac{m-1}{n}} d\theta \,\widehat{V}_1 (e^{i\theta_m})
\right \rangle_{\text{disk}} \ .
\end{equation}
This integral can be rewritten as
\begin{equation}
  \widehat{W}(\Psi_L,\mathcal{V})
 = -\frac{1}{2\pi i}\sum_{n = 1}^\infty \left\langle
  \mathcal{V}(0)\, c(1) \left\{ \exp\left( -\int_0^{2\pi} d\theta\, \widehat{V}_1(e^{i\theta}) \right) - 1\right\}
\right \rangle_{\text{disk}} \ .
\end{equation}
Hence, at least for this particular $\Phi$, we find a similar result to the bosonic case,
\begin{equation}
  \widehat{W}(\Phi,\mathcal{V}) = \mathcal{A}_{\Phi}^{\text{disk}}(\mathcal{V})
  - \mathcal{A}_{0}^\text{disk}(\mathcal{V}) \ .
\end{equation}

Inserting R-R-vertex operators on the disk is somewhat more subtle as
one has to pick the vertex operators in an asymmetric picture
\cite{Bianchi:1991eu,Polchinski:1987tu,DiVecchia:1997pr,Billo:1998vr,DiVecchia:1999rh}.
Moreover, to preserve the arguments made above, it is necessary to
pick a representation of the vertex operator which has total
$\phi$-momentum $-2$ and doesn't have any additional insertions of the
$\xi$-ghost zero-mode besides the factor of $(\xi +\tilde{\xi})$ that
will be inserted by hand.  The advantage of such a representation is
that it allows us to drop the second term in $\mathcal{O}_L$ as we did
in the NS-NS case. Such representations exist, but contain an infinite
number of terms \cite{Billo:1998vr}:
\begin{equation}
  \mathcal{V} = (\xi+\tilde{\xi}) \sum_{M = 0}^\infty \mathcal{V}^{(M)} (k,z,\tilde{z}) \ ,
\end{equation}
where
\begin{equation}
   \mathcal{V}^{(M)}(z,\bar z) = a_M \Omega_{AB} 
  \mathbb{V}^A_{-1/2 + M} 
  \tilde{\mathbb{V}}^B_{-3/2 -M} (\bar z) \ ,
\end{equation}
and the $a_M$ are constants, $\Omega_{AB}$ is a spinor
representation of the R-R-field of interest and
\begin{align}
  \mathbb{V}^A_{-1/2+M}(z) &= \partial^{M-1} \eta(z) \ldots \eta(z) 
   c(z) S^A(z) e^{(-\frac{1}{2} +M) \phi(z)} e^{i k X(z)/2} \ ,
\\
  \mathbb{V}^A_{-1/2+M}(z) &= \bar{\partial}^{M} \tilde{\xi}(\bar z) \ldots \bar \partial \tilde{\xi}(\bar z) 
  \tilde{c}(\bar{z}) \tilde{S}^A(\bar{z}) e^{(-\frac{3}{2} -M) \tilde{\phi}(\bar{z})} e^{i k \tilde{X}(\bar{z})/2} \ .
\end{align}
Noting that each term has one more $\xi$ than $\eta$ and a factor of
$e^{(-\frac{1}{2} +M)\phi + (-\frac{3}{2}-M)\tilde\phi}$, which
saturates the $\phi$-momentum of the disk, we can, as in the NS-NS
case, drop the second term in $\mathcal{O}_L$ given in (\ref{susyO})
from the computation and the same results follow.  Note that we are
free to pick other representations of the NS-NS vertex.  This choice
is convenient only in that it simplifies the relationship between
$\widehat{W}(\Phi,\mathcal{V})$ and the closed string one-point
function.  See also \cite{Michishita:2004rx} for a computation of the R-R invariants without
using this more complicated vertex operator.

Given that one can compute the R-R one-point function, the reader will
immediately wonder if it is possible to compute the R-R charges of a
given background.  Here we offer a few general remarks.  We leave a detailed analysis to future work.
In general, computing the R-R charges using $\widehat W(\Phi,\mathcal{V})$ is difficult because of the on-shell
constraint on the R-R vertex operator.  The on-shell constraint
typically allows one only to compute the coupling of the zero-mode of
the R-R field to the brane, which gives something proportional to the
integral of the R-R charge over the brane world volume (including the
infinite volume factor for the brane world-volume).  For the special
case of the D-instanton, there are no volume factors and the zero-mode
of the R-R tadpole is proportional to the number of D-instantons.

Even in the D-instanton case, however, this is not a manifestly topological quantity.  It
is only for classical solutions $\Phi$ that we can interpret
$\widehat{W}(\Phi,\mathcal{V})$ as being a closed string one-point function.
For example, since $\widehat W(\Phi,\mathcal{V})$ is linear in $\Phi$, if we allow $\Phi$ to
be an arbitrary state, there is no way that $\widehat W(\Phi,\mathcal{V})$
could always be an integer.  It appears, then, that
$\widehat{W}(\Phi,\mathcal{V})$ cannot be used to classify different $\Phi$'s as
having different charges off-shell.

\section*{Acknowledgments}
We would like to that A. Awad, S. Das, W. Merrel, Y. Okawa, and
B. Zwiebach for useful discussions and A. Hashimoto and M. Schnabl for comments on the draft.  We would also like to thank the participants of the {\em String field theory and related aspects} workshop for many useful comments.  This work was supported by
Department of Energy Grant No. DE-FG01-00ER45832.

\appendix

\section{Computation of the surface term}\label{SurfaceTermComp}
In this appendix, we explain the steps between (\ref{TwoPointFull}) and (\ref{TwoPointFullSurface}). 

Define the adjoint action of $\Psi$ by
\begin{equation}
  \text{ad}_\Psi A = \Psi* A - (-1)^{\text{gh}(A)} A*\Psi \ .
\end{equation}
Note that because of the grading,
\begin{equation}
  \left(\ad_\Psi\right)^2 A = \ad_{\Psi^2} A \ .
\end{equation}
We also have
\begin{equation}
 \{Q_B,\ad_\Psi\} = \ad_{Q_B \Psi} = -\ad_{\Psi^2} \ ,
\end{equation}
where in the last step we use that $\Psi$ satisfies the classical equations of motion.
Now, consider the two-point function with $\mathcal{V}_1 = \{Q_B,\mathcal{O}(i)\}$,
\begin{equation}
  \mathcal{A} = \sum_{n = 0}^\infty \left(\prod_{i = 1}^{n+1} \int_0^{\infty} dT_i \right)
  \langle \mathcal{I}| \{Q_B,\mathcal{O}(i)\}
  b_0 D_{T_1} \left(\prod_{i = 2}^{n+1} \ad_{\Psi} b_0 D_{T_i}\right) \mathcal{V}_2 (i) |\mathcal{I}\rangle \ .
\end{equation}
Impose a short distance cutoff on the length of the propagator,
\begin{equation}
  \mathcal{A}_\epsilon =  \sum_{n = 0}^\infty \int_{\epsilon/2}^{\infty} dT \left(\prod_{i = 1}^{n+1} \int_0^{\infty} dT_i \right)
  \delta(\mbox{$\sum_i$} T_i - T)
  \langle \mathcal{I}| \{Q_B,\mathcal{O}(i)\}
  b_0 D_{T_1} \left(  \prod_{i = 2}^{n+1} \ad_{\Psi} b_0 D_{T_i}\right) \mathcal{V}_2 (i) |\mathcal{I}\rangle \ .
\end{equation}
Now, push the $Q_B$ to the right:
\begin{multline}
 \mathcal{A}_\epsilon =  \sum_{n = 0}^\infty \int_{\epsilon/2}^{\infty} dT \left(\prod_{i = 1}^{n+1} \int_0^{\infty} dT_i \right)
  \delta(\mbox{$\sum_i$} T_i - T)
  \\
  \biggl\{
 - \langle \mathcal{I}| \mathcal{O}(i)
  \partial_{T_1} D_{T_1} \left(\prod_{i = 2}^{n+1}  \ad_{\Psi} b_0 D_{T_i}\right) \mathcal{V}_2 (i) |\mathcal{I}\rangle
  \\
  -
  \sum_{m = 1}^n \langle \mathcal{I}| \mathcal{O}(i)
  b_0 D_{T_1} \left(\prod_{i = 2}^{m}  \ad_{\Psi} b_0 D_{T_{i}}\right) \left(  \ad_{\Psi^2} b_0 D_{T_{m+1}}\right) \left( \prod_{i = m+2}^{n+1} \ad_{\Psi} b_0 D_{T_i}\right) \mathcal{V}_2 (i) |\mathcal{I}\rangle
  \\
-
   \sum_{m = 1}^n\langle \mathcal{I}| \mathcal{O}(i)
  b_0 D_{T_1} \left(\prod_{i = 2}^{m}   \ad_{\Psi} b_0 D_{T_{i}} \right)\left(  \ad_{\Psi} \partial_{T_i} D_{T_{m+1}}\right) \left( \prod_{i = m+2}^{n+1} \ad_{\Psi} b_0 D_{T_i}\right) \mathcal{V}_2 (i) |\mathcal{I}\rangle
  \biggr\} \ .
\end{multline}
Note that some of the terms have derivatives on the moduli.  Integrating by parts, these derivatives can be made to act on the delta-function and interpreted as derivatives with respect to $T$.  We write
\begin{equation}
  \mathcal{A}_\epsilon = \mathcal{A}_1+\mathcal{A}_2 \ ,
\end{equation}
with $\mathcal{A}_1$ given by the terms where the derivatives hit the delta-function,
\begin{multline}
  \mathcal{A}_1
   = 
   -\sum_{n = 0}^\infty \int_{\epsilon/2}^{\infty} dT \frac{\partial}{\partial T}\left(\prod_{i = 1}^{n+1} \int_0^{\infty} dT_i \right)
   \delta(\mbox{$\sum_i$} T_i - T)
  \\
  \biggl\{
  \langle \mathcal{I}| \mathcal{O}(i) D_{T_1} \left(  \ad_{\Psi} b_0 D_{T_i}\right)^n \mathcal{V}_2 (i) |\mathcal{I}\rangle
  \\
+
   \sum_{m = 1}^n\langle \mathcal{I}| \mathcal{O}(i)
  b_0 D_{T_1} \left( \prod_{i = 2}^m \ad_{\Psi} b_0 D_{T_i}\right)\left(  \ad_{\Psi} D_{T_{m+1}}\right) \left( \prod_{i = m+2}^{n+1} \ad_{\Psi} b_0 D_{T_i}\right) \mathcal{V}_2 (i) |\mathcal{I}\rangle
  \biggr\} \ .
   \\
   =
    -\sum_{n = 0}^\infty \int_{\epsilon/2}^{\infty} dT \frac{\partial}{\partial T}\left(\prod_{i = 1}^{n+1} \int_0^{\infty} dT_i \right)
   \delta(\mbox{$\sum_i$} T_i - T)
  \biggl\{
  \langle \mathcal{I}| \mathcal{O}(i) D_{T_1} \left(\prod_{i = 2}^{n+1}  \{b_0,\ad_{\Psi}\} D_{T_i}\right) \mathcal{V}_2 (i) |\mathcal{I}\rangle \biggr\} \ ,
\end{multline}
and $\mathcal{A}_2$ the rest,
\begin{multline}
 \mathcal{A}_2 =  \sum_{n = 0}^\infty \int_{\epsilon/2}^{\infty} dT \left(\prod_{i = 1}^{n+1} \int_0^{\infty} dT_i \right)
  \delta(\mbox{$\sum_i$} T_i - T)
  \\
  \biggl\{
 - \langle \mathcal{I}| \mathcal{O}(i) \left(\prod_{i = 2}^{n+1}  \ad_{\Psi} b_0 D_{T_i}\right) \mathcal{V}_2 (i) |\mathcal{I}\rangle \qquad \qquad \qquad  \qquad \qquad \qquad \nonumber
 \end{multline}
 \begin{multline}
  -
  \sum_{m = 1}^n \langle \mathcal{I}| \mathcal{O}(i)
  b_0 D_{T_1} \left(\prod_{i = 2}^{m}  \ad_{\Psi} b_0 D_{T_i}\right) \left(  \ad_{\Psi^2} b_0 D_{T_{m+1}}\right) \left( \prod_{i = m+2}^{n+1} \ad_{\Psi} b_0 D_{T_i}\right) \mathcal{V}_2 (i) |\mathcal{I}\rangle
  \\
+
  \delta(T_{n+1}) \sum_{m = 1}^n\langle \mathcal{I}| \mathcal{O}(i)
  b_0 D_{T_1} \left(\prod_{i = 2}^{m}   \ad_{\Psi} b_0 D_{T_{i}} \right)\left(  \ad_{\Psi}\right) \left( \prod_{i = m+1}^{n} \ad_{\Psi} b_0 D_{T_i}\right) \mathcal{V}_2 (i) |\mathcal{I}\rangle
  \biggr\} \ .
\end{multline}
To simplify this, note that the first term in the $\{ \ \}$'s vanishes since
\begin{equation}
  \langle \mathcal{I} |\mathcal{O}(i) \ad_{\Psi} = \ad_{\Psi} \mathcal{V}_2(i) |\mathcal{I}\rangle = 0 \ .
\end{equation}
This also kills the third term when $m = n$. We are left with
\begin{multline}
 \mathcal{A}_2 =  \sum_{n = 0}^\infty \int_{\epsilon/2}^{\infty} dT \left(\prod_{i = 1}^{n+1} \int_0^{\infty} dT_i \right)
  \delta(\mbox{$\sum_i$} T_i - T) \sum_{m = 1}^n 
  \\
  \biggl\{
  -
   \langle \mathcal{I}| \mathcal{O}(i)
  b_0 D_{T_1} \left(\prod_{i = 2}^{m}  \ad_{\Psi} b_0 D_{T_i}\right) \left(  \ad_{\Psi^2} b_0 D_{T_{m+1}}\right) \left( \prod_{i = m+2}^{n+1} \ad_{\Psi} b_0 D_{T_i}\right) \mathcal{V}_2 (i) |\mathcal{I}\rangle
  \\
+
 \delta(T_{n+1})\langle \mathcal{I}| \mathcal{O}(i)
  b_0 D_{T_1} \left(\prod_{i = 2}^{m}   \ad_{\Psi} b_0 D_{T_i} \right)\left(  \ad_{\Psi^2} b_0 D_{T_{m+1}}\right)
  \left( \prod_{i = m+2}^{n} \ad_{\Psi} b_0 D_{T_i}\right) \mathcal{V}_2 (i) |\mathcal{I}\rangle
  \biggr\} \ .
\end{multline}
This vanishes since the second term in the $\{ \ \}$'s is zero for $n<2$, while the first term is zero for $n<1$.  If follows that
\begin{equation}
  \mathcal{A}_{\epsilon} = \mathcal{A}_1 \ ,
\end{equation}
from which (\ref{TwoPointFullSurface}) follows.

\section{Changing the height of a state by reparametrization}\label{reparamdiscussion}

\begin{figure}
\centerline{
\scalebox{.5}{\begin{picture}(685,141)(-30,-10)   
\includegraphics{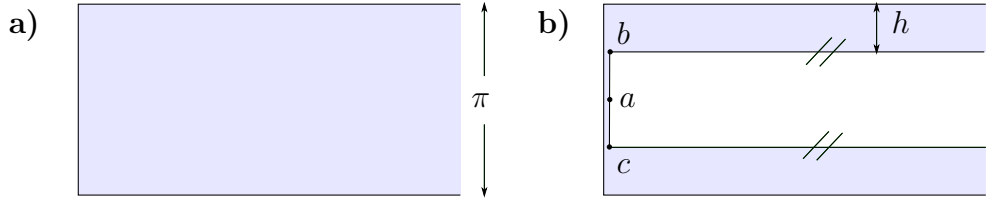}  
\end{picture}}
\begin{picture}(0,0)(342.5,0)
\put(160,39){$\pi$}
\put(216,39){$a$}
\put(215,63){$b$}
\put(215,14){$c$}
\put(319,67){$h$}
\put(-15,67){\bf a)}
\put(185,67){\bf b)}
\end{picture}
}
\caption{In a a typical state is shown with width $\pi/2$.  In b, a
modified state is shown which reduces to the identity state near the
midpoint.  The lines $ab$ and $ac$ are to be identified as well as the
lines extending to the right of $b$ and $c$ as shown with the hatches.
In the actual geometry of interest the thin vertical strip of
worldsheet to the left of $cab$ would be of zero thickness.
\label{Reparam}}
\end{figure}

In this appendix, we briefly discuss why the height of an insertion
$\Psi_{\text{cl}}$ may be changed by a reparametrization and, hence, a
gauge transformation.  In figure \ref{Reparam}a a state is shown in
strip coordinates.  To decrease the height of the insertion, we
replace the region of the state near the midpoint with the identity
state so that it has no effect when inserted into the propagator.  The
rest of the state is shrunk to a width $h$.  This is shown in figure
\ref{Reparam}b

The important point to recognize is that the height $h$ can be
adjusted by simply rescaling the identity and strip segments of the
state in a way that keeps the whole length of the state fixed.  For
example, if $0<\theta<\pi$ is a coordinate on the unit circle, we can
perform the reparametrization
\begin{equation}
  \tilde{\theta}(\theta) = \left\{ \begin{matrix}
  \rho \theta & \theta<h
  \\
  \vphantom{a} & 
  \\
  \frac{\pi}{2} - \frac{\pi - 2\rho h}{\pi-2h}(\pi/2- \theta) & h<\theta<\pi/2
   \end{matrix}  \right. \ ,
\end{equation}
where we also define $\tilde\theta(\pi - \theta) = \pi - \tilde
\theta(\theta)$.  This map scales $h \to \rho h$.  Note that because the
identity state is invariant under symmetric reparametrizations which
preserve the midpoint and endpoints, there is considerable flexibility
in the choice of $\tilde \theta(\theta)$ in the region $h<
\theta<\pi - h$.

Note also that picking $\rho = \pi/2h$ leads to a singular
reparametrization; the entire region $h<\theta<\pi - h$ is mapped to
the midpoint.  However, as long as the state is inserted into a larger
worldsheet geometry, this transformation remains smooth.  One may also
worry that $\tilde \theta(\theta)$ could create problems if there are
operators near the midpoint (points $b$ and $c$ in figure
\ref{Reparam}b).  Though we have no basis for doing so (as we do not
have a regularity condition on our string field), we assume that
operators insertions near the midpoint are sufficiently mild that this
will not be a problem.

\bibliography{c}\bibliographystyle{utphys}

\end{document}